\documentclass{ws-ijbc}

\usepackage{graphicx}
\usepackage{epstopdf}
\usepackage{hyperref} 

\begin{document}

\catchline{}{}{}{}{} 

\markboth{H.~T.~Moges et al.}{The phase space structure in a 3D galactic bar potential}


\title{The evolution of the phase space structure along pitchfork and period-doubling bifurcations in a 3D galactic bar potential}

\author{H.~T.~MOGES}
\address{Nonlinear Dynamics and Chaos Group, Department of Mathematics and Applied Mathematics,\\
 University of Cape Town, Rondebosch 7701, South Africa\\
\email{MGSHEN002@myuct.ac.za}}

\author{M.~KATSANIKAS}
\address{Research Center for Astronomy, Academy of Athens,\\
Soranou Efessiou 4, GR-11527 Athens, Greece\\
\vspace{5pt}
School of Mathematics, University of Bristol, Fry Building, \\Woodland Road, Bristol, BS8 1UG, United Kingdom.\\
\email{mkatsan@academyofathens.gr}}

\author{P.~A.~PATSIS}
\address{Research Center for Astronomy, Academy of Athens,\\
 Soranou Efessiou 4, GR-11527 Athens, Greece\\
\email{patsis@academyofathens.gr}}

\author{M.~HILLEBRAND}
\address{Max Planck Institute for the Physics of Complex Systems, \\ 
N\"{o}thnitzer Stra\ss e 38, 01187 Dresden, Germany\\
\vspace{5pt}
Center for Systems Biology Dresden,\\
Pfotenhauer Stra\ss e 108, 01307 Dresden, Germany\\
\email{hillebrand@pks.mpg.de}}

\author{Ch.~SKOKOS}
\address{Nonlinear Dynamics and Chaos Group, Department of Mathematics and Applied Mathematics,\\ 
University of Cape Town, Rondebosch 7701, South Africa\\
\email{haris.skokos@uct.ac.za ; haris.skokos@gmail.com}}

\maketitle

\begin{history}
Accepted for Publication
\end{history}

\begin{abstract}
We investigate how the phase space structure of a three-dimensional (3D) autonomous Hamiltonian system evolves across a series of successive two-dimensional (2D) and 3D pitchfork and period-doubling bifurcations, as the transition of the parent families of periodic orbits (POs) from stability to simple instability leads to the creation of new stable POs. Our research illustrates the consecutive alterations in the phase space structure near POs as the stability of the main family of POs changes. This process gives rise to new families of POs within the system, either maintaining the same or exhibiting higher multiplicity compared to their parent families. Tracking such a phase space transformation is challenging in a 3D system. By utilizing the color and rotation technique to visualize the four-dimensional (4D) Poincar\'e surfaces of section of the system, i.e.~projecting them onto a 3D subspace and employing color to represent the fourth dimension, we can identify distinct structural patterns. Perturbations of parent and bifurcating stable POs result in the creation of tori characterized by a smooth color variation on their surface. Furthermore, perturbations of simple unstable parent POs beyond the bifurcation point which lead to the birth of new stable families of POs, result in the formation of figure-8 structures of smooth color variations. These figure-8 formations surround well-shaped tori around the bifurcated stable POs, losing their well-defined forms for energies further away from the bifurcation point. We also observe that even slight perturbations of highly unstable POs create a cloud of mixed color points, which rapidly move away from the location of the PO. Our study introduces, for the first time, a systematic visualization of 4D surfaces of section within the vicinity of higher multiplicity  POs. It elucidates how, in these cases, the coexistence of regular and chaotic orbits contributes to shaping the phase space landscape.
\end{abstract}
\keywords{Chaos and dynamical systems; galactic dynamics; 4D surfaces of section; pitchfork bifurcation; period-doubling bifurcation.}



\section{Introduction} 
\label{sec:introduction}

In this paper, we study the evolution of the phase space structure in a three-dimensional (3D) barred galactic potential before and after successive two-dimensional (2D) and 3D pitchfork and period-doubling bifurcations of families belonging to the ``x$1$-tree" \citep{skokos2002orbital}. The x$1$ family is the main family of periodic orbits (POs), which provides the building blocks for the bar in 2D galactic models \citep{contopoulos1980orbits,athanassoula1983orbits}. In general, the POs of the x$1$ family have an elliptical-like morphology in non-axisymmetric galactic systems with their major axis extending along the major axis of the bar. As the system's energy increases, the x$1$ POs may develop cusps or even two loops at their edges. The x$1$ family undergoes a series of bifurcations, which in 3D systems, builds the so-called x$1$-tree \citep{skokos2002orbital,SPA02b}. We note that a pitchfork bifurcation\footnote{A pitchfork bifurcation is also called ``direct'' or ``supercritical''  (see e.g.~\citep[Sect.~2.4.3]{contopoulos2002order} and \citep[Sect.~3.4]{strogatz2018nonlinear})} of a family of a stable PO leads to the creation of a pair of new stable POs of equal period, with the parent family becoming unstable. On the other hand, in the case of a period-doubling bifurcation the destabilization of the parent family produces a new stable PO having twice the period of the original one (see e.g.~\cite[Sect.~7.1b]{LL92}). 

The phase space structure before and beyond a 3D  pitchfork bifurcation of single members of the x$1$ family has been studied in \citep{katsanikas2011structure}. Extending and in some sense completing that paper, we study in the present work how the phase space of a 3D system evolves during successive pitchfork bifurcations, also including in our analysis period-doubling bifurcations. In particular, our investigation focuses on the evolution of the phase space structure within the following scenario: Initially, a 2D pitchfork bifurcation of the x$1$ family creates a new 2D family. Subsequently, this new 2D family undergoes a bifurcation, transitioning into a 3D family via the same pitchfork mechanism. Our analysis then delves into the alterations observed in the system's phase space as the 3D family experiences two consecutive 3D period-doubling bifurcations. 

The first two cases are 2D and 3D bifurcations that occur when a stable family of POs becomes simple unstable (see Sect.~\ref{subsec:StablilityofPOs} for the definition of the various instability kinds of POs). At this transition, we have the emergence of two new stable families of POs having the same period as the parent one. The other two cases refer to 3D bifurcations, which also occur when a stable 3D family of POs becomes simple unstable. However, in this instance, a new 3D family emerges whose multiplicity is double that of the parent family.

The direct visualization of the phase space of 3D Hamiltonian systems is not feasible because the Poincar\'e surface of section (PSS) of these systems are four-dimensional (4D). Starting with Froeschl\'e's pioneering work \citep{froeschle1970numerical}, several methods attempting the visualization of the 4D PSSs have been proposed and implemented. Some frequently applied techniques include the 2D and 3D projections  of the 4D PSS \citep{martinet1981invariant, vrahatis1997structure} and its stereoscopic views \citep{martinet1981number}, along with the method of plotting low-dimensional slices of the whole PSS \citep{froeschle1970numerical, froeschle1972numerical,LROBK14,richter2014visualization,OLKB16}. 

Besides visualization per se, another fundamental obstacle when studying 4D sections of the phase space is to trace the zones of influence of the various periodic orbits and thus assess their contribution to the formation of  specific phase space structures  in each investigated case. In other words, we always have difficulty defining the ``neighborhood" of the periodic orbit. Contrarily to the by-eye appreciation of the borders of a stability  island, of the border of a chaotic layer between two stability islands, or even of the presence of sticky zones in a 2D PSS, the landscape of phase space structure in a 4D space is not discernible. For instance, a minor perturbation in the initial conditions of an unstable PO might lead to an orbit in a chaotic region of the phase space, while a larger one could unexpectedly place the perturbed orbit on a torus surrounding a nearby stable PO, resulting in regular behavior. Therefore, we must proceed with utmost care when we consider perturbations of POs.

In our study, we use the color and rotation technique \citep{patsis1994using} to visualize the  4D PSS and to understand the underlying dynamics of our 3D galactic-type Hamiltonian system. Additional applications of this technique can be found in \citep{katsanikas2011structure,katsanikas2011structure1,zachilas2013structure,patsis2014phasea,
patsis2014phaseb,agaoglou2021visualizing,katsanikas2022phase}.

Another notable complexity in our study arises from the necessity to visualize not just the immediate vicinity of a particular PO of single multiplicity in the phase space, but to also consider  cases where double structures emerge (such as in cases of period-doubling). Additionally, we want to incorporate within our visualization the modifications that arise in the phase space around the parent family, which is now unstable. Given the close proximity of both the parent and newly generated POs in phase space, our goal is to depict the phase space volume encompassing both neighborhoods. This endeavor has been  attempted only once in the past \citep{katsanikas2011structure1} and needs to be investigated in more detail.

The paper is organized as follows: The Hamiltonian model for the description of the rotating barred galaxy is presented in Sect.~\ref{sec:model}. The stability of POs and the color and rotation technique are discussed in Sect.~\ref{sec:NumericalMethods}. The phase space structure before and after the considered bifurcations is investigated in Sect.~\ref{sec:result}. Finally, the main findings and conclusions of our work are summarized in Sect.~\ref{sec:Discussion and Conclusion}.

\section{The 3D Hamiltonian model} 
\label{sec:model}

The Hamiltonian function governing the 3D rotating model of a barred galaxy is given by
\begin{equation}
H = \dfrac{1}{2} (p_x^2 + p_y^2 + p_z^2) + V(x, y, z) - \Omega_b(xp_y - yp_x),
\label{eq:3DBarredGAlaxy} 
\end{equation}
where $x$, $y$ and $z$ are  the Cartesian coordinates and $p_x$, $p_y$ and $p_z$ are the canonically conjugate momenta of a test particle, $\Omega_b$ represents the pattern speed of the bar and $V$ is the total potential, which, in our case, consists of three components: A disk, a bar and a bulge. $H$ is the Jacobi constant of the model,  and we will also refer to its numerical value, $E_j$, as the ``energy''. 

To be more specific, $V$ comprises the sum of three individual potential terms delineating an axisymmetric part, $V_0 = V_{\textit{sphere}} + V_{\textit{disc}}$, where $V_{\textit{sphere}}$ and $V_{\textit{disc}}$ are the bulge and disk terms respectively, alongside a component representing the galaxy's bar, $V_{\textit{bar}}$.

The potential of the bulge is given by a Plummer sphere \citep{plummer1911problem} 
\begin{equation}
\label{eq:Plummer}
    V_{\textit{sphere}}(x,y,z) = -\dfrac{GM_S}{\sqrt{x^2+y^2+z^2+\epsilon^2_s}},
\end{equation}
where $\epsilon_s$ is the scale length of the bulge, $M_S$ is its mass, and $G$ is the gravitational constant. 

The disk is represented by a Miyamoto–Nagai disc \citep{miyamoto1975three} 
\begin{equation}
\label{eq:MiyamotoNagai}
    V_{\textit{disc}}(x,y,z) = -\dfrac{GM_D}{\sqrt{x^2+ y^2 + (A + \sqrt{B^2+z^2})^2}},
\end{equation}
where $M_D$ is the mass of the disk, while $A$ and  $B$ are respectively horizontal and vertical scale lengths. 

The bar component of the potential is 
\begin{equation}
\label{eq:Vbar}
     V_{\textit{bar}} = -\pi Gabc \dfrac{\rho_c}{3} \int_{\lambda}^{\infty}  \dfrac{du}{\Delta(u)} (1-m^2(u))^{3}, 
\end{equation}
where $\rho_c = \dfrac{105}{32\pi}\dfrac{GM_B}{abc}$, $m^2(u) = \dfrac{y^2}{a^2+u} + \dfrac{x^2}{b^2+u} + \dfrac{z^2}{c^2+u}$, $\Delta^2(u) = (a^2+u)(b^2+u)(c^2+u)$, $\lambda$ is the unique positive solution to $m^2(\lambda) = 1$ for the regions outside the bar ($m \ge 1$) and $\lambda = 0$ for the region inside the bar ($m < 1$), and $a > b > c$ are the semi-axes of the bar ~\citep{pfenniger19843d}. The corresponding mass density is
\begin{equation}
    \label{eq:Bar_mass_dens}
    \rho = \left\{ \begin{array}{lll} \displaystyle \rho_c (1-m^2)^2 &
    \mbox{for} &  m \le 1,  \\
    & & \\
    0 & \mbox{for} &  m > 1, \\
    \end{array} \right.
\end{equation}
with $m^2 = \dfrac{y^2}{a^2} + \dfrac{x^2}{b^2} + \dfrac{z^2}{c^2}$, i.e., the major axis of the bar is along the y-axis (see also \citep{pfenniger19843d,skokos2002orbital}). 

In our study, we use the following values for the model's parameters: $\Omega_b=0.054$,  $a=6$, $b=1.5$, $c=0.6$, $A=3$, $B=1$, $\epsilon_s=0.4$, $M_B=0.2$, $M_D=0.72$, where the units are as follows: 1 kpc (distance), 1 Myr (time), $2\times 10^{11}$ solar masses (\(M_\odot\)) (mass) and the total mass $G(M_S + M_D + M_B) = 1$. This arrangement corresponds to what was named ``model D'' in \citep{SPA02b}.

\section{Numerical techniques} 
\label{sec:NumericalMethods}

In our study, we numerically determine the stability of POs and implement the method of color and rotation to visualize the system's 4D PSS. We briefly present these techniques below.

\subsection{Stability of periodic orbits} 
\label{subsec:StablilityofPOs}

The tracing of POs is done by using a Newton iterative method with an accuracy of at least $10^{-10}$. To find the linear stability of POs in 3D Hamiltonian systems we follow the algorithms introduced by \citep{broucke1969stability} and \citep{hadjidemetriou1975continuation}. The stability of a PO is determined by the eigenvalues and eigenvectors of the so-called Floquet (monodromy) matrix $\mathbf{M}(T)$, which is the fundamental solution matrix of the PO's variational equations evaluated at a time equal to one period, $T$, of the orbit. The various stability kinds are classified according to the stability indices $b_1$, $b_2$ and a quantity typically denoted by $\Delta$, which are calculated from the coefficients of the characteristic polynomial of $\mathbf{M}(T)$ (see e.g.~\citep{contopoulos1985simple} and \citep{S01} for more details about the meaning and the computation of $b_1$, $b_2$ and  $\Delta$). A PO is called stable ``S" if $|b_1|, |b_2| < 2$, and  $\Delta > 0$, simple unstable ``U" if $|b_1|  < 2, |b_2| > 2$, or if $|b_1| > 2, |b_2| < 2$, and  $\Delta > 0$, and double unstable ``DU" if $|b_1|, |b_2| > 2$ and $\Delta > 0$. If $\Delta < 0$ the PO is complex unstable ``$\Delta$", a case that is encountered but not investigated in the present study.

\subsection{The method of color and rotation} 
\label{subsec:ColorandRotation}

Exploring the phase space dynamics of 2D Hamiltonian systems is straightforward using their 2D PSS, and the same applies to 2D symplectic maps since their phase portrait is also 2D. However, this approach becomes impractical when dealing with systems of higher dimensions, in which the PSS or phase portrait dimension is larger than two. A method proposed to gain insight into the structure of the 4D PSS in 3D Hamiltonian systems, or the phase space of 4D symplectic maps, is known as the method of color and rotation~\citep{patsis1994using} (more details about this method can also be found  in~\citep{katsanikas2011structure}). In our study, the color and rotation method is implemented by following these steps:

\begin{enumerate}
	\item For a given initial condition, we numerically integrate the system's equations of motion obtaining the orbit's evolution in time. We find many intersections of this orbit with the  PSS defined by $y = 0$, $p_y>0$. As a result, we get sets of points in the 4D $(x, p_x, z, p_z)$ space.
	\item We choose different 3D subspaces of the 4D PSS, such as $(x, p_x, z)$ or $(x, z, p_z)$, and plot the 3-tuples of the orbit's points.
	\item We color each point in the 3D subspace according to the value of its fourth coordinate.
        \item Additionally, we rotate the 3D objects using some plotting software to enhance our comprehension of the developed structures.
	 
\end{enumerate}

When the color and rotation technique is used on a 4D PSS, the presence of a torus with a smooth color variation on its surface indicates the presence of regular orbital behaviour. On the other hand, the chaotic nature of an orbit  is reflected in the irregular behavior of the points in the 3D subspace and/or by the mixing of these points' colors representing the fourth dimension \citep{patsis1994using,katsanikas2011structure,katsanikas2013instabilities}.

The system's equations of motion for a fixed $E_j$ value and for an initial condition $(x_0, p_{x_0}, z_0, p_{z_0})$ on the PSS $y=0$ (with the  $p_{y_0} >0$  defined by the given $E_j$ value) are numerically solved by implementing the fourth-order Runge-Kutta integration method. We use an appropriately chosen integration time step for each orbit studied in our work, which guarantees that the relative energy error remains less than $10^{-10}$ for the orbit's whole time evolution.

\section{Numerical results} 
\label{sec:result}

\subsection{Characteristic and stability diagram}
In what follows, we study in detail the evolution of the phase space structure before and after a series of 2D and 3D bifurcations of POs in the dynamical system defined by the Hamiltonian \eqref{eq:3DBarredGAlaxy}. Specifically, we investigate the alteration in the 4D PSS, introduced by a 2D and a 3D pitchfork bifurcation, alongside two 3D period-doubling bifurcations. We note that in a 2D bifurcation, the POs of both the parent and the created families are planar, lying on the $(x,y)$ symmetry plane of the galaxy model. 

\begin{figure*}[t]
	\centering
	\includegraphics[width=0.85\textwidth,keepaspectratio]{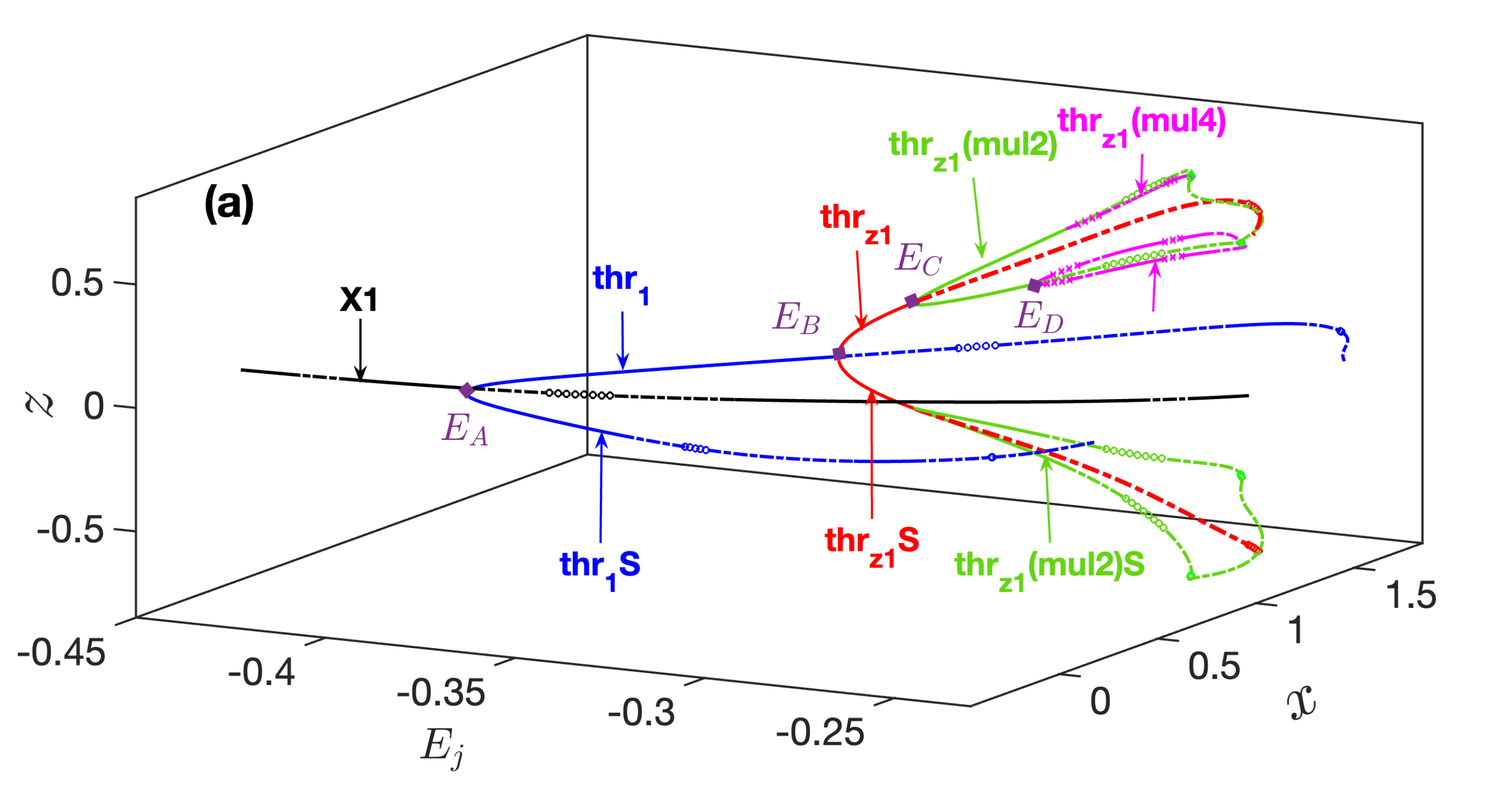}
	\includegraphics[width=0.85\textwidth,keepaspectratio]{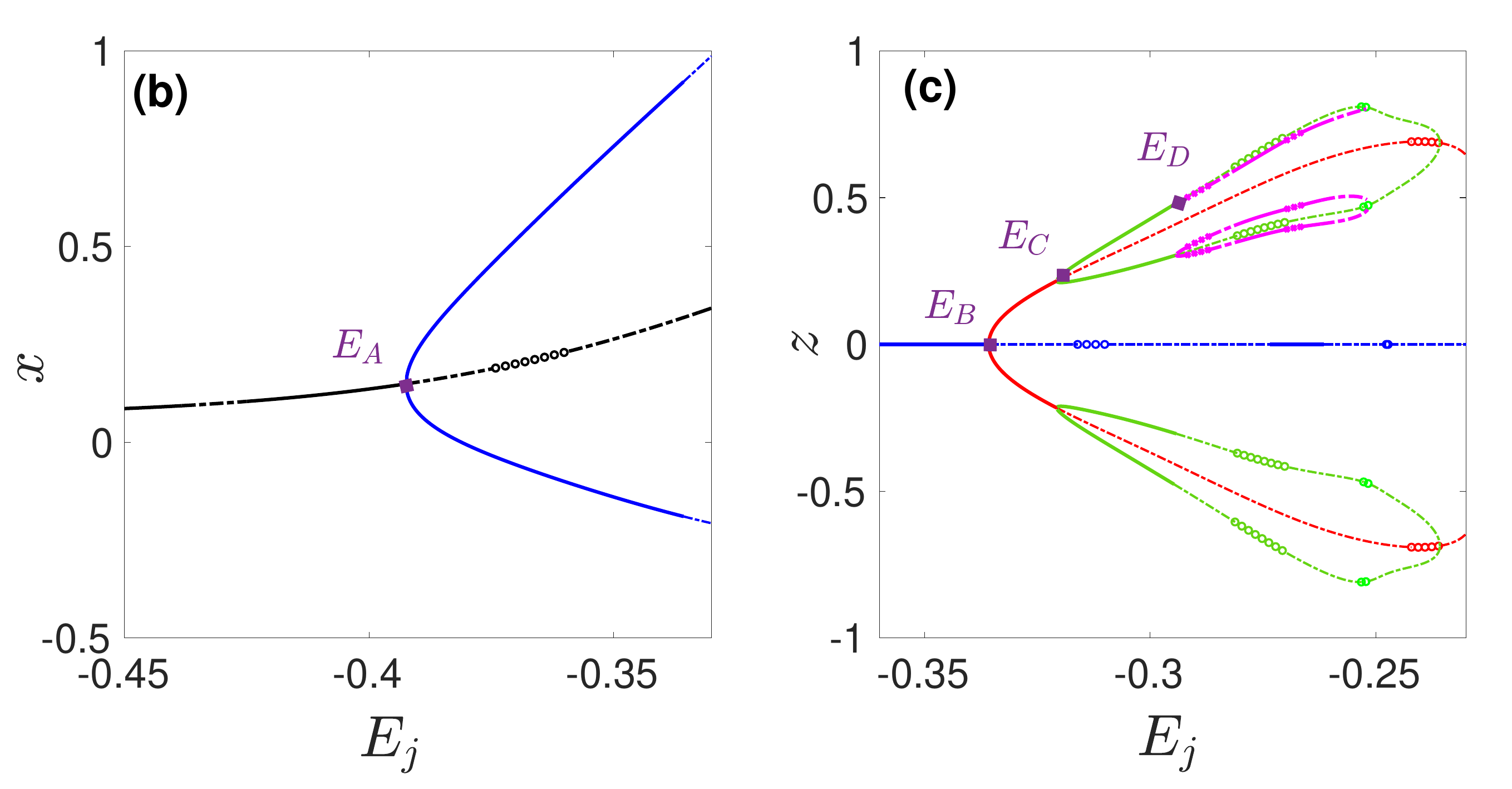}	
	\caption{(a) The 3D characteristic diagram $(E_j,x,z)$ of the families of POs we consider in our study: thr$_1$ (blue curves), thr$_{z1}$ (red curves), thr$_{z1}$(mul2) (green curves), thr$_{z1}$(mul4) (magenta curves) and their symmetric counterparts  thr$_1$S, thr$_{z1}$S, thr$_{z1}$(mul2)S, thr$_{z1}$(mul4)S, as well as of the main 2D family x$1$ (black curves). (b) The 2D characteristic diagram $(E_j,x)$ of the 2D families shown in (a), obtained by a 2D projection of part of panel (a). (c) The 2D characteristic diagram $(E_j,z)$ of part of the diagram shown in (a). The different stability types of the POs are denoted by different curve styles: stable (solid curves), simple unstable (dashed curves),  double unstable (open circles) and complex unstable (``$\times$'' symbols). The values of the system's energy $E_j$ where bifurcations are taking place are $E_A=-0.3924$,  $E_B=-0.3356$,  $E_C=-0.3203$ and $E_D=-0.2943$. We note that the thr$_{z1}$(mul4) has four branches in panels (a) and (c), but the upper two branches (for $z>0$) practically overlap.}
	\label{fig:f01}
\end{figure*}

As depicted in Figs.~\ref{fig:f01}(a) and \ref{fig:f01}(b), showcasing the ``characteristic diagram" (i.e., the coordinates of the initial conditions of the periodic orbits as a function of $E_j$) for the main planar family x$1$ (represented by black curves), this family experiences a bifurcation at energy $E_A=-0.3924$. This bifurcation leads to the birth of two new planar (2D) families, which are initially stable (blue curves in Fig.~\ref{fig:f01}). The first family is named ``thr$_1$'' and its symmetric with respect to the y-axis counterpart is denoted as ``thr$_1$S'' \footnote{The name indicates a three-to-one resonant family, referring to resonances encountered on deferentially rotating galactic disks (see e.g.~\citep[Sect.~3.4]{contopoulos2002order}).}.

Subsequently, the family thr$_1$ becomes vertically unstable and undergoes another bifurcation at energy $E_B=-0.3356$,  resulting in the appearance of two new, initially stable, 3D families, denoted by ``thr$_{z1}$'' and ``thr$_{z1}$S'' [red curves in Figs.~\ref{fig:f01}(a) and \ref{fig:f01}(c)]. Again, here ``S" indicates symmetry, this time with respect to the equatorial plane of the model. We note that the family thr$_1$S undergoes a similar bifurcation at $E_B$, which we nevertheless do not include in the diagrams of Fig.~\ref{fig:f01} for the sake of clarity. 

At energy $E_C=-0.3203$, the thr$_{z1}$ family undergoes a period-doubling bifurcation, giving rise to a family of multiplicity two, named ``thr$_{z1}$(mul2)'', together with its symmetric family, ``thr$_{z1}$(mul2)S'', bifurcating from the thr$_{z1}$S family at the same energy. These two families are depicted in green in Figs.~\ref{fig:f01}(a) and \ref{fig:f01}(c). We note that the multiplicity of a PO refers to the number of times it intersects the PSS (where $y = 0$ and $p_y > 0$) within a single orbital period $T$. 

Then, at energy  $E_D=-0.2943$ the thr$_{z1}$(mul2) family [and its counterpart thr$_{z1}$(mul2)S] undergoes another period-doubling bifurcation becoming simple unstable, generating at the same time the stable ``thr$_{z1}$(mul4)'' family [and its counterpart ``thr$_{z1}$(mul4)S''] of multiplicity four [magenta curves in  Figs.~\ref{fig:f01}(a) and \ref{fig:f01}(c), which are also indicated by  arrows in  Fig.~\ref{fig:f01}(a)]. We note that the upper magenta curve of thr$_{z1}$(mul4) in Figs.~\ref{fig:f01}(a) and \ref{fig:f01}(c) actually corresponds to two curves, which practically overlap at the presented projections. The different stability kinds of the POs of each family are represented by different curve styles: stable parts of families of POs are represented by solid curves, simple unstable POs by dotted curves, double unstable POs by open circles and complex unstable POs by ``$\times$'' symbols.

\begin{figure*}[t]
	\centering
  	\includegraphics[width=\textwidth,keepaspectratio]{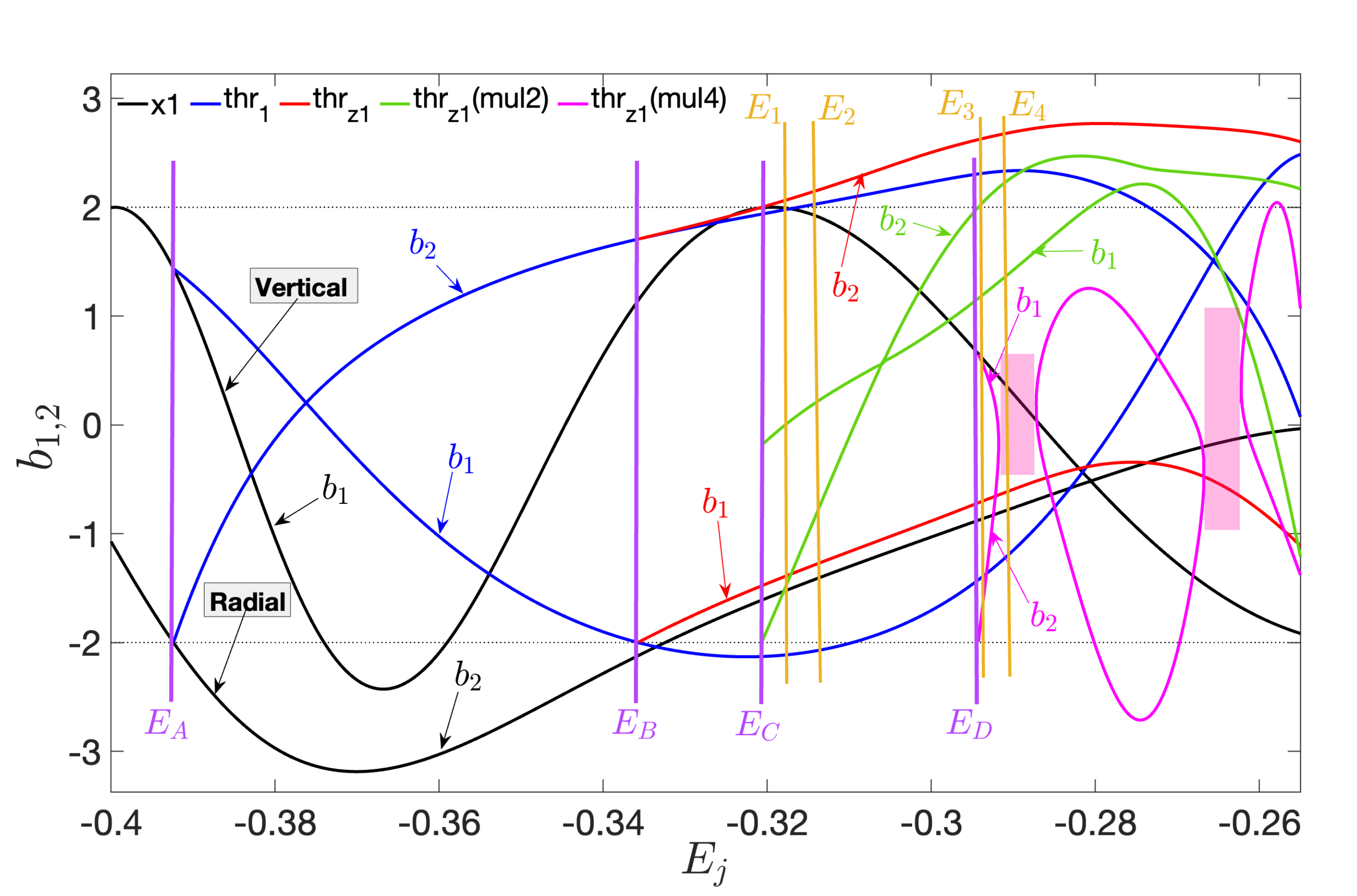}
	\caption{Stability diagram, i.e.~the vertical ($b_1$) and the radial ($b_2$) stability indices as a function of the energy E$_j$, of the considered families of POs, namely x$1$ (black curves), thr$_{1}$ (blue curves), thr$_{z1}$ (red curves), thr$_{z1}$(mul2) (green curves), and thr$_{z1}$(mul4) (magenta curves) (see also the legend at the top of the figure). The four vertical purple lines denoted by $E_A$, $E_B$,  $E_C$ and $E_D$ respectively indicate the bifurcation points at which the creation of family thr$_{1}$ from x$1$, thr$_{z1}$ from thr$_{1}$,  thr$_{z1}$(mul2) from thr$_{z1}$, and thr$_{z1}$(mul4) from thr$_{z1}$(mul2) takes place. The magenta-shaded areas denote energy intervals where the thr$_{z1}$(mul4) family is complex unstable and thus $b_1,b_2$ are not defined. The four vertical orange  lines indicate the energies $E_1=-0.3183$, $E_2=-0.3157$, $E_3=-0.2941$ and $E_4=-0.2907$ of the four specific cases discussed in Sect.~\ref{Sec:(mul2)thrz1} (see also Table \ref{tab:tab1}).}
\label{fig:f02}
\end{figure*}

In Fig.~\ref{fig:f02}, we give the so-called ``stability diagram'' \citep{contopoulos1985simple}, i.e.~the evolution of the stability indices $b_1$ and $b_2$ of the various families of POs we consider in our study, as a function of $E_j$. The stability indices $b_1$ and $b_2$ are respectively related to vertical and planar (radial) perturbations. Both indices of each family are plotted with the same color, but different colors are used for different families of POs. Black curves correspond to the main family x$1$. The transition of x$1$ from stability to simple instability at energy $E_A$ results in the creation of family thr$_1$ and its symmetric counterpart, thr$_1$S (blue curves). This is the first bifurcation we study in detail in this work, and it happens when the radial index $b_2$  of x$1$ crosses the $b=-2$ axis. 

The thr$_1$ (and thr$_1$S)  family remains stable until the transition to instability at energy $E_B$, where we have the birth of the 3D stable families thr$_{z1}$ and thr$_{z1}$S (red curves). This bifurcation takes place when the vertical index $b_1$ of thr$_1$ crosses the $b=-2$ axis. 

At even larger energies the 3D families thr$_{z1}$(mul2) and thr$_{z1}$(mul2)S (green curves) are introduced at the first transition from stability to simple instability of the thr$_{z1}$ family, which is caused by the crossing, in this case, of the $b_2$ index of thr$_{z1}$ with the $b=2$ line at energy $E_C$. 

Then, at $E_D$, through another period-doubling bifurcation, as the green $b_2$ index crosses the $b=2$ axis, the multiplicity four family thr$_{z1}$(mul4) (magenta curves) is born from thr$_{z1}$(mul2). We note that the values of the stability indices for each of the families, thr$_1$, thr$_{z1}$, thr$_{z1}$(mul2) and thr$_{z1}$(mul4) are respectively equal to the ones of their symmetric counterparts thr$_1$S, thr$_{z1}$S,  thr$_{z1}$(mul2)S and thr$_{z1}$(mul4)S. As we can observe in Fig.~\ref{fig:f02}, the stability curves of the families that are introduced in the system by period-doubling do not emerge from the intersection of a stability index of the parent family with the $b=2$ axis. This would have happened only if we had computed the stability indices of the parent family by considering it to be  of double multiplicity (see Appendix A in \citep{SPA02b} and \citep{patsis2019orbital}).

\begin{figure}[t]
	\centering
	\includegraphics[width=0.85\textwidth,keepaspectratio]{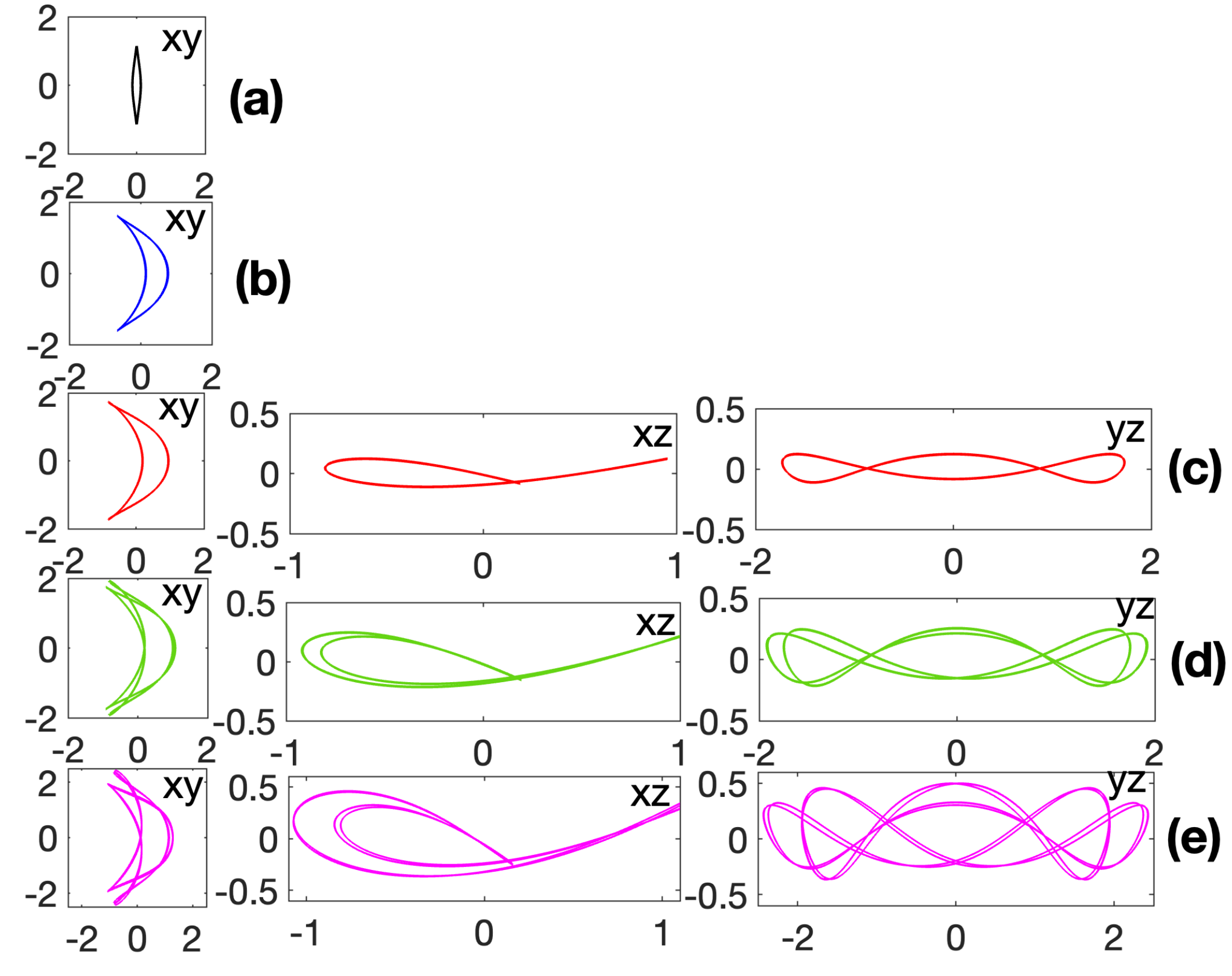}
 \caption{Representative stable POs of the  five considered families: (a) an elliptical-like x$1$ PO for $E_j = -0.41$, (b) a thr$_1$ PO  for $E_j = -0.35$, (c) a thr$_{z1}$ PO  for  $E_j = -0.3306$,  (d) a multiplicity two thr$_{z1}$(mul2)  PO for $E_j = -0.3183$ and (e) a multiplicity four thr$_{z1}$(mul4) PO for $E_j = -0.2831$.  From left to right in (c), (d), and (e), the orbits are shown at different 2D projections, namely the $(x,y)$, $(x,z)$ and $(y,z)$ planes.}
\label{fig:f03}
\end{figure}

As we can see from the plots of Fig.~\ref{fig:f03}, where representative stable POs of families x$1$ [Fig.~\ref{fig:f03}(a)], thr$_1$ [Fig.~\ref{fig:f03}(b)], thr$_{z1}$ [Fig.~\ref{fig:f03}(c)],  thr$_{z1}$(mul2) [Fig.~\ref{fig:f03}(d)] and thr$_{z1}$(mul4) [Fig.~\ref{fig:f03}(e)] are presented, the POs of x$1$, thr$_1$ and thr$_{z1}$ have multiplicity one, since they cross the $y=0$ axis with $p_y>0$ only once in one orbital period. On the other hand, the PO of thr$_{z1}$(mul2) [Fig.~\ref{fig:f03}(d)] has two such crossing in one period and consequently its multiplicity is two, while the  thr$_{z1}$(mul4) PO [Fig.~\ref{fig:f03}(e)] is of multiplicity four. 

The characteristic diagram in Fig.~\ref{fig:f01} and the stability diagram in Fig.~\ref{fig:f02}, together with the morphology of the POs of each family depicted in Fig.~\ref{fig:f03}, will be our primary guides in our effort to follow the dynamical evolution of phase space structures we find in the 4D PSS, as the various bifurcations take place.

\subsection{A 2D  pitchfork bifurcation} 
\label{Sec:``8"}

Fig.~\ref{fig:f02} indicates that the main planar x$1$ family is stable for small $E_j$ values, as both its stability indices (black curves) are between $-2$ and $2$. As the system's energy $E_j$ increases, the x$1$ family becomes simple unstable at $ E_A= -0.3924$, leading to the creation of two, initially stable, 2D families of POs: thr$_1$ [a representative PO of which is shown in Fig.~\ref{fig:f03}(b)], and its symmetric counterpart thr$_1$S. The thr$_1$ (and thr$_1$S) family is stable for $E_A < E_j < E_B= -0.3356$ (blue curves in Fig.~\ref{fig:f02}), whereas the x$1$ family remains simple unstable in this interval, apart from the interval $-0.374 < E_j < -0.359$  in which it is double unstable.

\begin{figure*}[t]
	\centering
 	\includegraphics[width=0.85\textwidth,keepaspectratio]{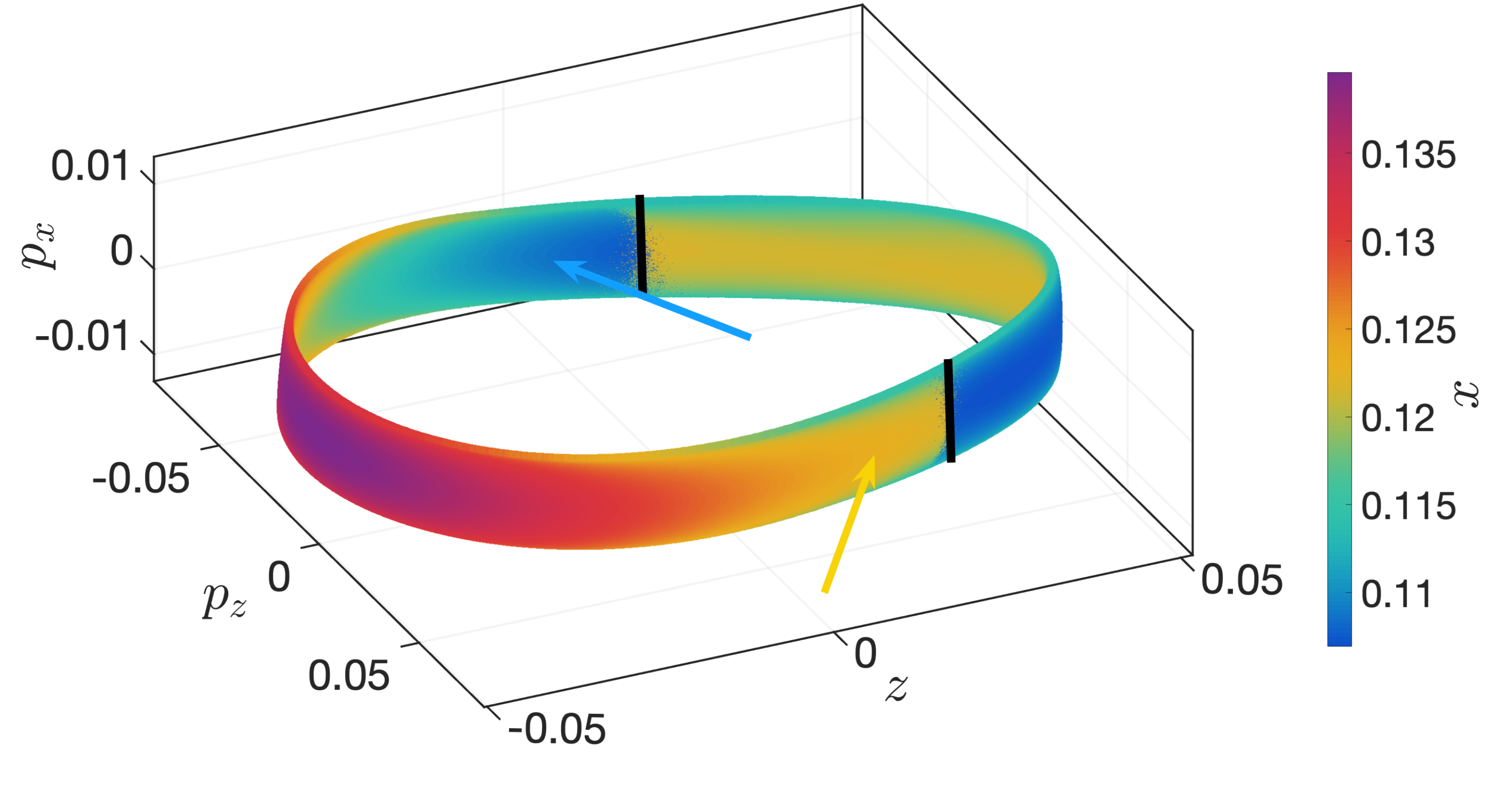}
	\caption{The 3D projection  $(p_x, z, p_z)$ of the 4D PSS of a quasi-periodic orbit forming a torus around the stable x$1$ PO at $E_j = -0.41 <E_A$. The color scale at the right part of the figure is used to color the consequents according to the value of their $x$ coordinate. The initial condition of the shown orbit is obtained by a $\Delta z =  5 \times 10^{-2}$ perturbation of the PO's initial condition in the $z$ direction. The primary hue of the color shifts from one side of the torus to the other (from the inner to the outer regions and vice versa) once it moves beyond the areas demarcated by bold, black straight-line segments (see text).}
 \label{fig:f04}
\end{figure*}

A representative example of the system's 4D PSS in the vicinity of the POs of the main family x$1$, for energies before the critical point of the 2D pitchfork bifurcation at $E_A$, and in particular for $E_j=-0.41 < E_A$, is presented in Fig.~\ref{fig:f04}. At this energy, the x$1$ PO is surrounded by invariant tori because it is stable. The particular torus shown in Fig.~\ref{fig:f04} is obtained by perturbing the x$1$ PO's initial condition $z_0$ by $\Delta z =  5 \times 10^{-2}$. The existence of an invariant torus around the stable x$1$ PO in the 3D subspace $(p_x, z, p_z)$ of the 4D PSS shown in Fig.~\ref{fig:f04}, together with the color variation along this object, indicate that the orbits obtained by small perturbations of the x$1$ PO remain confined around it, exhibiting regular motion. The smooth color variation observed on the torus in Fig.~\ref{fig:f04} implies a similar consistent configuration of this object in the fourth dimension of the PSS, namely in $x$. This suggests that projecting this orbit into any possible 3D subspace of the PSS would generate a torus-like object displaying an ordered variation of colors.

We note that in Fig.~\ref{fig:f04}, moving anticlockwise from the point indicated by an orange arrow at the exterior surface of the torus  (clockwise from the one pointed out by a blue arrow  on the interior surface of the torus) we follow points which retain their color by switching from the external to the internal side of the torus (and vice versa) when they reach the transition region between the orange and blue colors (the regions of transition are indicated by heavy solid, black lines in Fig.~\ref{fig:f04}). Such smooth color transitions on a projected torus were discussed in detail in \citep{katsanikas2011structure}. Upon examining the color variation closely, we notice it is complex. The variation occurs in both the toroidal and poloidal directions. The bold, black, straight line segments outlined in Fig.~\ref{fig:f04} signify the area beyond which essentially the observation of a color hue along the same side of the torus ceases and continues along the other side. We underline that the regular character of an orbit is associated with the smoothness of the color variation and not with the exact pattern it follows.

When $E_j$ becomes slightly greater than $E_A$, the POs of the x$1$ family are simple unstable, while the planar 2D POs of the bifurcated thr$_{1}$ family, as well as the POs of its symmetric family thr$_{1}$S, are stable. The simple instability of the x$1$ family within this energy range is a radial instability. This is indicated by the fact that the $b_2$ index, which is related to  radial perturbations, crosses the $b = -2$ axis becoming $b_2 < -2$. Regarding vertical perturbations, the $b_1$ index of the $x1$ family remains within the $-2<b_1<2$ range, indicating vertical stability.

Perturbations of the unstable x$1$ POs in the radial direction, for example by changing their $x$ coordinate by $\Delta x$, results in 2D, planar orbits and in the creation of a typical  ``figure-8'' structure on the $(x, p_x)$ projection of the 4D PSS (note that in this case $z=p_z=0$), with the unstable x$1$ PO located at the center of the ``8" formation, while the stability islands of the two, newly bifurcated stable families are positioned within the two lobes of the figure-8 structure (see e.~g.~\cite[Fig.~2.45]{contopoulos2002order}). Actually, this arrangement represents the typical configuration observed in 2D PSSs under such circumstances.

\begin{figure}[t]
	\centering
 	\includegraphics[width=0.85\textwidth,keepaspectratio]{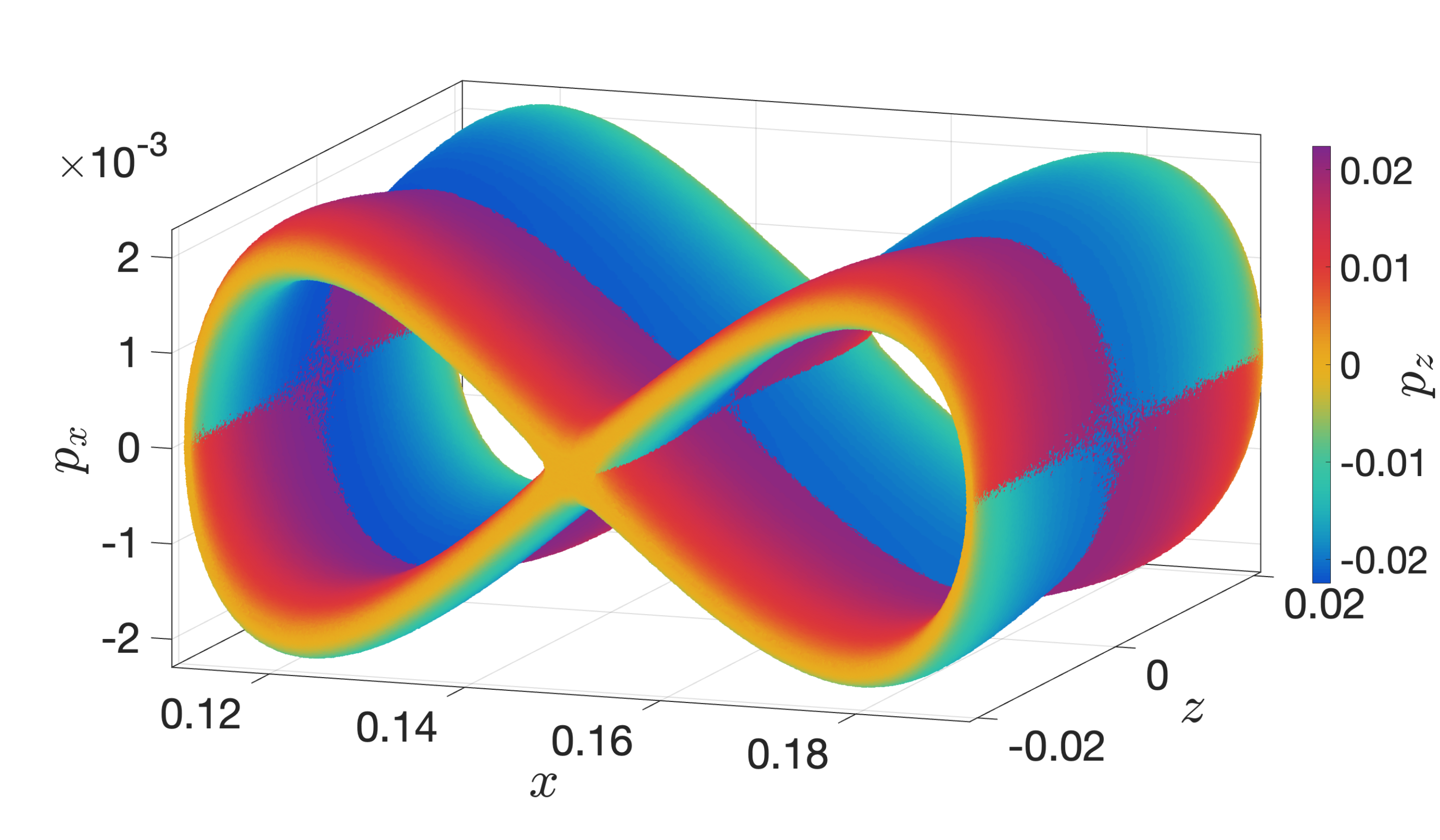}
	\caption{The 3D colored $(x, p_x, z)$ projection of the system's 4D PSS for a perturbation of the unstable x$1$ PO by $\Delta z =  2 \times 10^{-2}$ for $ E_j = -0.3919$ ($E_A < E_j < E_B$). }
 \label{fig:f05}
\end{figure}

However, an interesting feature of our 3D system, is that the figure-8 configuration is also retained in the $(x,p_x)$ projection of the PSS for a range of, rather small, vertical perturbations of the unstable x$1$ POs. For example, we found that for $E_j=-0.3919$ this structure is present for $\Delta z \leq 2\times 10^{-2}$. In Fig.~\ref{fig:f05} we present the 3D colored $(x, p_x, z)$ projection of the 4D PSS for a perturbation by $\Delta z =  2 \times 10^{-2}$ along the $z$ axis of the simple unstable x$1$ PO  for $ E_j = -0.3919> E_A$. Each point in this plot is colored according to its $p_z$ value. As we can observe, the $(x, p_x)$ projection of the created structure has the anticipated 8-shaped form with the unstable x$1$ PO located at the center of the projected ``8"  figure. It is worth noting that the consequents do not disperse in the $z$ direction; instead, they form a ribbon-like structure in the 4D PSS, something which is indicated by the points' smooth color variation. Similar to the torus of the quasi-periodic orbit in Fig.~\ref{fig:f04}, we also observe characteristic color transitions from the outer to the inner side of the structure for the current orbit. In Appendix A we discuss some practical aspects, which should be taken into account for studies of 3D  orbits near simple unstable POs that maintain vertical stability. To the best of our knowledge, the presentation of this 8-shaped  structure is the first  investigation of  phase space structures linked to $z$-axis perturbations of a radial unstable and vertically stable PO.  Our exploration is concentrated on phase space structures in the vicinity of this PO, which are extended beyond the 2D plane. Such interesting behaviors warrants further systematic investigation. 

\begin{figure}[t]
	\centering
	 \includegraphics[width=0.85\textwidth,keepaspectratio]{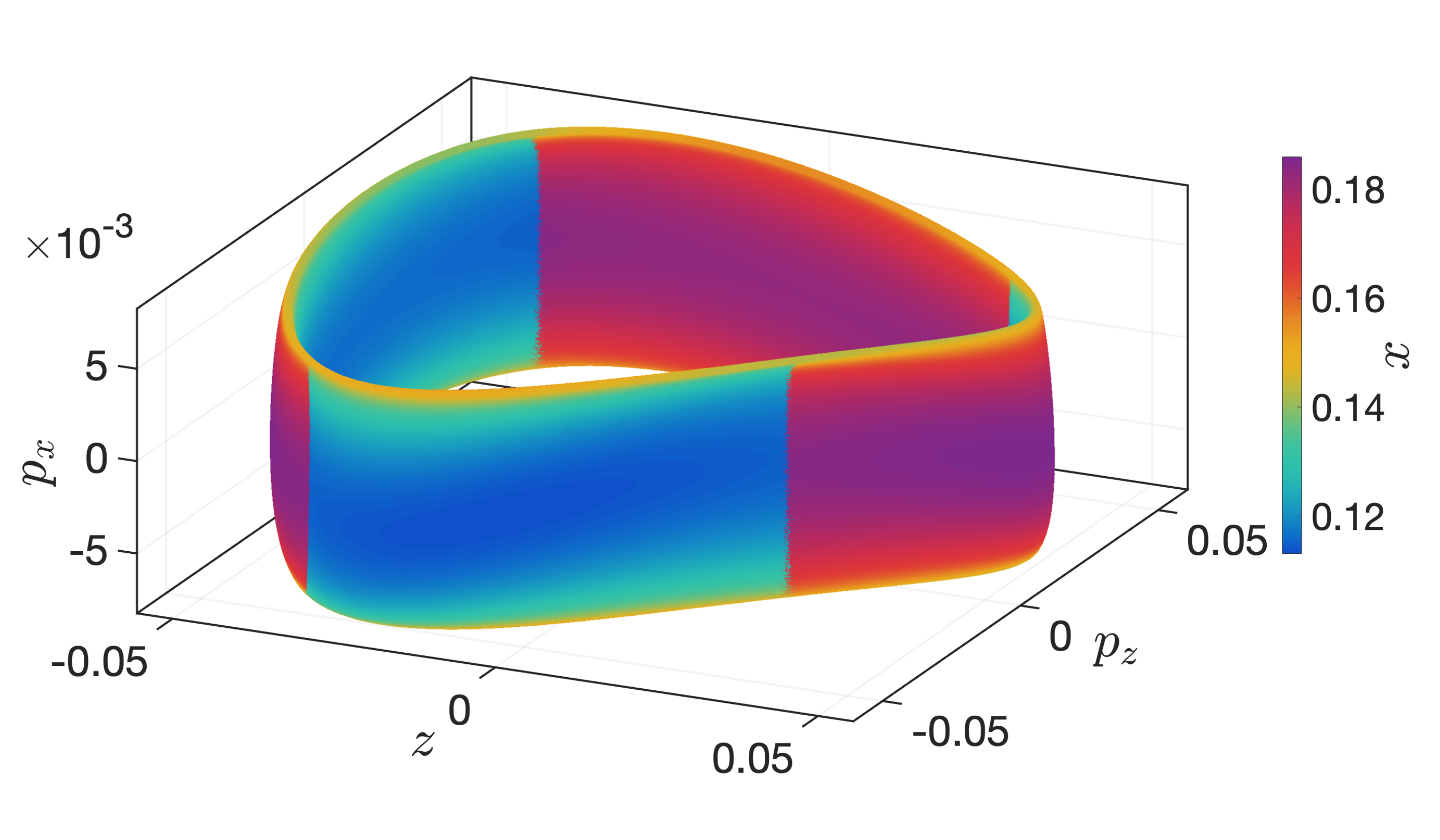}
	\caption{The 3D colored  $(p_x, z,  p_z)$  projection of the system's 4D PSS for a perturbation of the stable thr$_{1}$ PO by $\Delta z =  5 \times 10^{-2}$ for $ E_j = -0.3919$.} 
 \label{fig:f06}
\end{figure}

As expected, the perturbation of the stable planar thr$_{1}$ PO in the $z$ direction (Fig.~\ref{fig:f06}) results in the creation of an invariant torus around the PO, which is characterized by a smooth color variation, similar to what is seen in Fig.~\ref{fig:f04}. Once more, we note the transitions of distinct color zones (blue and red in this case) moving from the outer to the inner surface of the torus and vice versa, which occur along the regions where the darker blue- and red-colored areas intersect. Tori similar to the ones around the thr$_1$ POs are found around the thr$_1$S POs as well. These tori correspond to quasi-periodic orbits around the stable, bifurcated families and they are situated within the lobes of the 8-shaped structure illustrated, for instance, in Fig.~\ref{fig:f05} (see also the cases described in Sect.~\ref{Sec:thrz1}).

\subsection{A 3D  pitchfork bifurcation} 
\label{Sec:thrz1}

As the energy increases, the stability of the new families varies. A transition of the planar family thr$_1$ from stability to simple instability at energy $E_B= -0.3356$ creates two new, initially stable, 3D families of POs through a pitchfork bifurcation, i.e.~thr$_{z1}$ and thr$_{z1}$S (see Fig.~\ref{fig:f01}). Since there is no difference whether we follow thr$_{z1}$ or thr$_{z1}$S, our presentation focuses on describing the evolution of the phase space structure in the vicinity of thr$_{z1}$ orbits. The morphology of the thr$_{z1}$ family is seen in Fig.~\ref{fig:f03}(c), where the stable member of this family for $E_j = -0.3306$ is depicted. As we can see in Fig.~\ref{fig:f02}, the thr$_{z1}$ family becomes simple unstable at energy $E_C=-0.3203$ and further retains that kind of instability. Observing the stability indices of the thr$_1$ family in Fig.~\ref{fig:f02} (blue curves), we notice that it is its  vertical index, $b_1$, that intersects the $b=-2$ axis. Consequently, for $E_j>E_B$ the family thr$_1$ initially becomes vertically unstable, although it maintains its radial stability, as indicated by the evolution of its  $b_2$ index. 

\begin{figure*}[t]
	\centering
 	\includegraphics[width=0.9\textwidth,keepaspectratio]{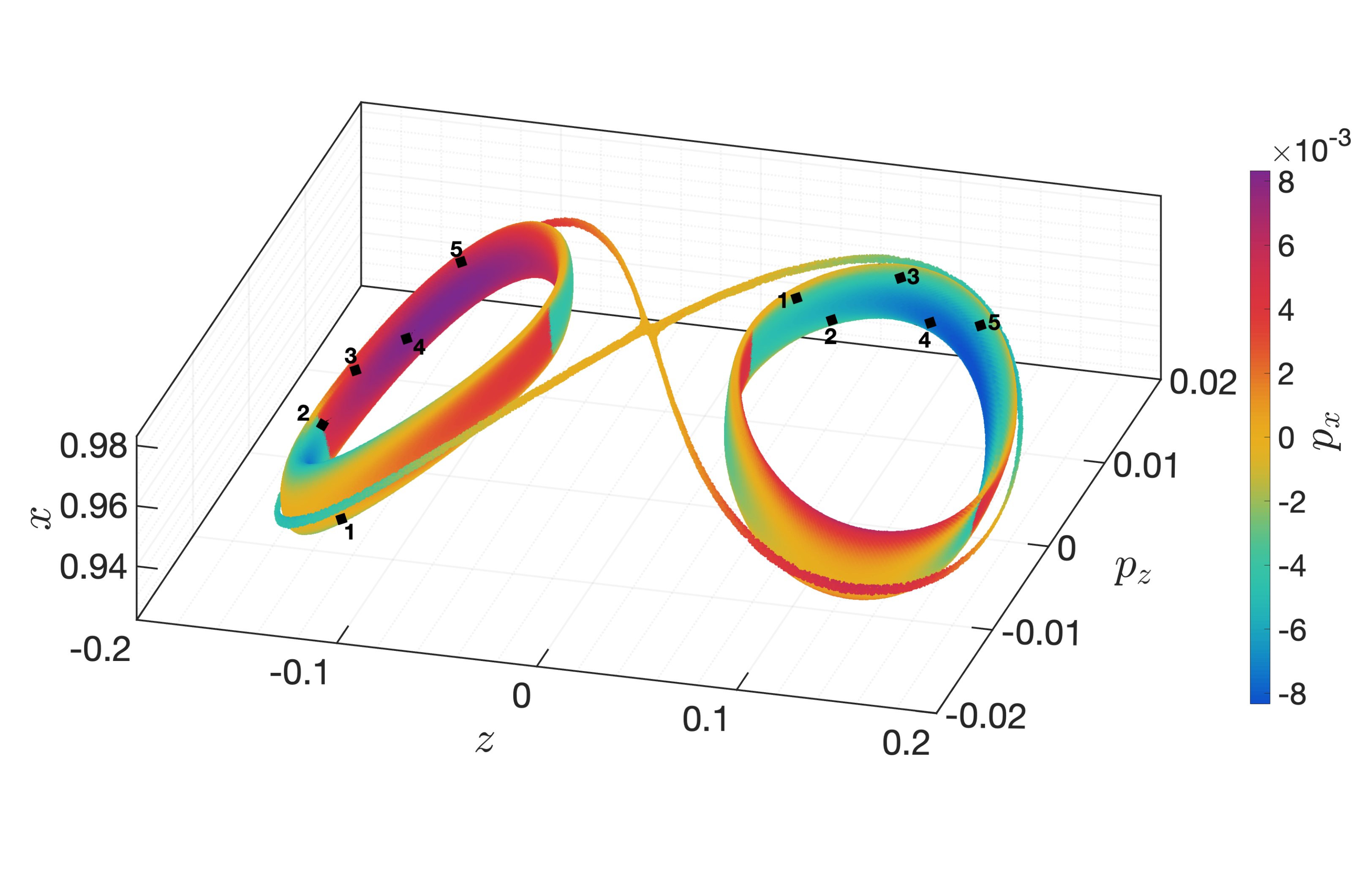}
	\caption{The 3D colored $(x,z,p_z)$ projection of the system's 4D PSS at $E_j = -0.3307$ for perturbations of the  simple unstable thr$_1$ PO by $\Delta z = 10^{-5}$ (figure-8 structure), as well as the stable thr$_{z1}$ (right torus)  and  thr$_{z1}$S (left torus) POs by $\Delta z =  5 \times 10^{-2}$. The two sequences of numbered points denoted by black diamonds show the first five consequents of orbits with initial conditions on each torus.}
 \label{fig:f07}
\end{figure*}

Let us first investigate the phase space structure for energies slightly larger than $E_B$ so that the planar thr$_1$ family is simple unstable and the 3D family thr$_{z1}$ (and thr$_{z1}$S) is stable. In  Fig.~\ref{fig:f07} we show the 3D projection $(x,z,p_z)$ of the system's PSS, for perturbations of the thr$_1$, thr$_{z1}$ and thr$_{z1}$S POs at $E_j = -0.3307$, with points being colored according to their $p_x$ values. The consequents of the orbit generated by a $\Delta z = 10^{-5}$ perturbation of the simple unstable 2D thr$_1$ PO create a thin figure-8 structure in the $(x, z, p_z)$ projection in Fig.~\ref{fig:f07}. Due to the unstable nature of the thr$_1$ PO, even a very small $\Delta z$ perturbation of the orbit in the $z$ direction results in chaotic orbits moving away from the PO located at the intersection of the two halves of the figure-8 structure. 

The perturbation in the $z$ direction of the two stable thr$_{z1}$  and  thr$_{z1}$S POs by $\Delta z =  5 \times 10^{-2}$ leads to quasi-periodic motion on the two invariant tori shown in Fig.~\ref{fig:f07}. The right torus corresponds to the perturbation of the thr$_{z1}$ PO, and the left one to the perturbation of the thr$_{z1}$S PO. Since the two bifurcated families are of single multiplicity, akin to the parent one, thr$_{1}$, the two tori are not connected to each other. Consequents of any initial condition on the right or left torus always remain on the same torus. The first five consequents of the two orbits depicted in Fig.~\ref{fig:f07} are denoted by numbered black diamond symbols. The 4D morphology of each one of the two tori in  Fig.~\ref{fig:f07} is similar to the one observed for the perturbations of the stable x$1$ (Fig.~\ref{fig:f04}) and thr$_1$ (Fig.~\ref{fig:f06}) POs. 

\begin{figure*}[t]
	\centering
 	\includegraphics[width=0.75\textwidth,keepaspectratio]{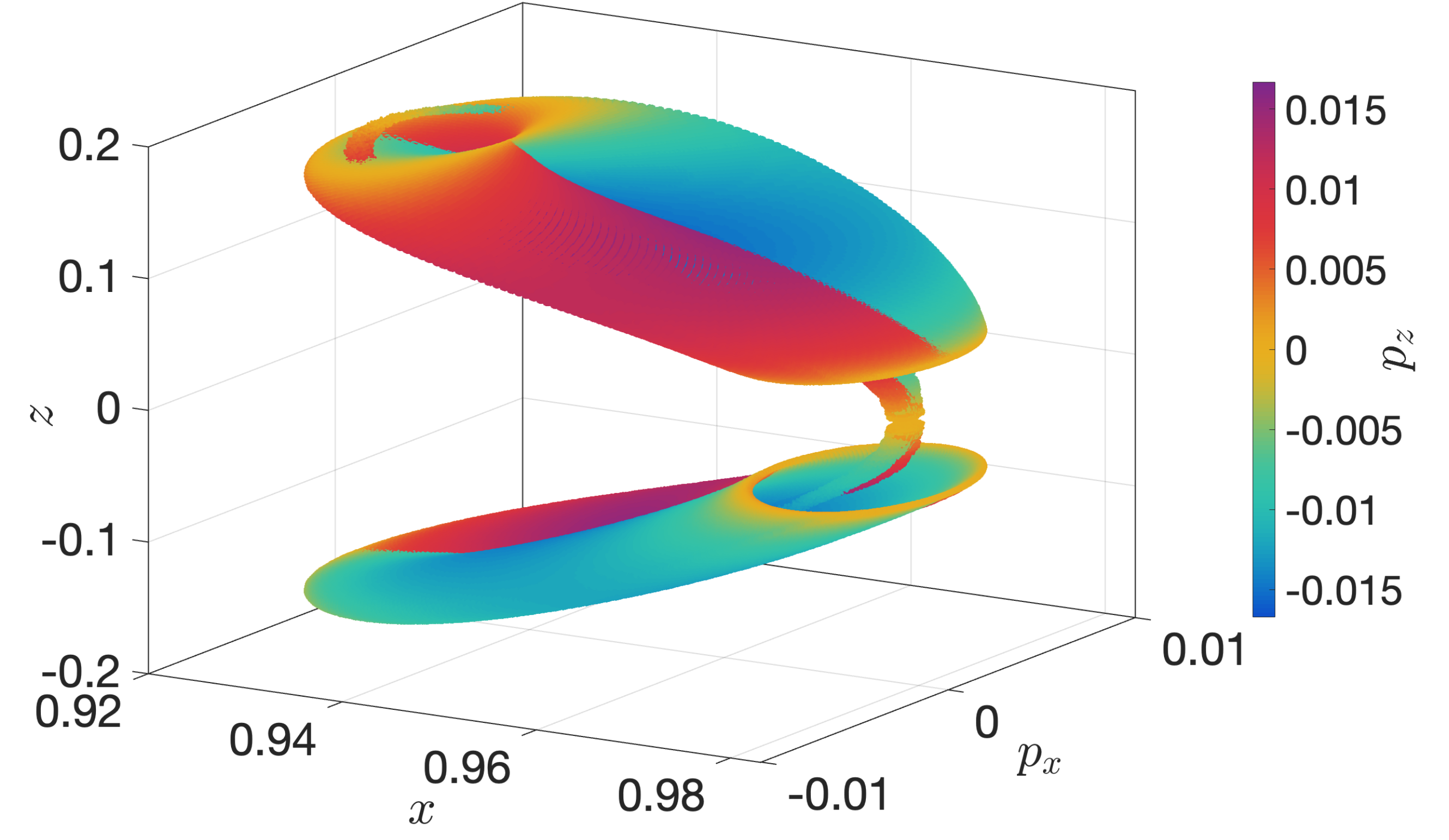}
	\caption{The same orbits depicted in Fig.~\ref{fig:f07}, but using the $(x,p_x,z)$ 3D projection.}
	 \label{fig:f08}
\end{figure*}

It is interesting to observe that these three orbits exhibit a different configuration in an alternate projection, although the perturbation of the thr$_1$ PO maintains the 8-shaped outline, while the quasi-periodic orbits  near the stable thr$_{z1}$ and thr$_{z1}$S POs retain their toroidal structure. In Fig.~\ref{fig:f08} we depict the same orbits as in Fig.~\ref{fig:f07}, but by using the $(x,p_x,z)$ projection of the PSS and the $p_z$ coordinate for giving the color to the consequents. In this case, we observe that the tori, corresponding to the  quasi-periodic orbits, around the stable periodic orbits of the newly bifurcated families surround an 8-shaped ribbon-like structure, which again has at the intersection of its lobes the unstable thr$_1$ PO.

\begin{figure*}[t]
	\centering
 	\includegraphics[width=0.8\textwidth,keepaspectratio]{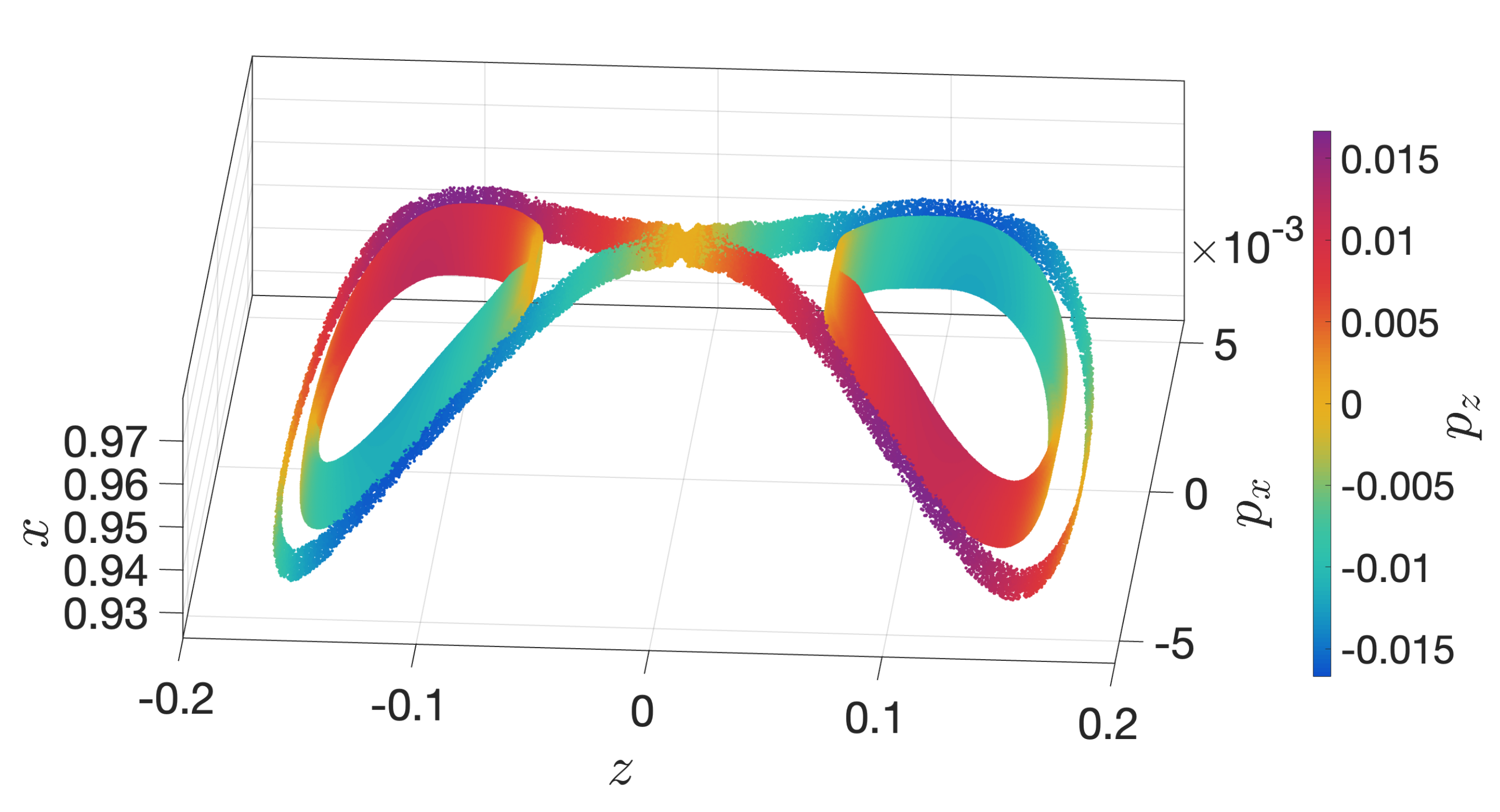}
	\caption{The $(x,p_x,z)$ 3D projection of a perturbed thr$_1$ PO by $\Delta z = 10^{-5}$, together with the tori corresponding to the quasi-periodic orbits around the POs thr$_{z1}$ and thr$_{z1}$S obtained by a $\Delta p_x = 2.5 \times 10^{-3}$ and $\Delta p_z = 1.2 \times 10^{-2}$ perturbation of each family. All orbits have $E_j = -0.3307$.}
	 \label{fig:f09}
\end{figure*}

It is worth noting that the relative projection of the tori and the 8-shaped structures depends on the chosen 3D projection and the perturbations we apply to the POs. For the same POs considered in Figs.~\ref{fig:f07} and \ref{fig:f08} we  apply a different set of perturbations, namely a $\Delta z = 10^{-5}$ perturbation to the simple unstable thr$_1$ PO, and both $\Delta p_x = 2.5 \times 10^{-3}$ and $\Delta p_z = 1.2 \times 10^{-2}$ to its stable bifurcations thr$_{z1}$ and thr$_{z1}$S, to obtain the configuration observed in Fig.~\ref{fig:f09}. We also note that the consequents creating the  8-shaped structure in Fig.~\ref{fig:f09}  start diffusing in the 4D PSS at time $t \approx 10^5$ and will occupy a larger volume of phase space abruptly, after integrating the orbit for time $t \gtrsim 5 \times 10^5$ (note that the orbits in Fig.~\ref{fig:f09} were integrated until $t=10^4$). It should be emphasized that the distinctive 8-shaped structures are observed when perturbations in $\Delta z$ are introduced. If we introduce relatively small radial perturbations, i.e.~$\Delta x$ or $\Delta p_x$, to the initial conditions of the planar POs of the thr$_1$ family, the resulting orbits will also remain planar. Then, we will observe in the $(x, p_x)$ projection of the 4D PSS closed curves around the simple unstable PO, resembling the patterns seen around the x$1$ family in Fig.~4 of  \citep{patsis2014phasea}.

\begin{figure*}[t]
	\centering
	\includegraphics[width=0.8\textwidth,keepaspectratio]{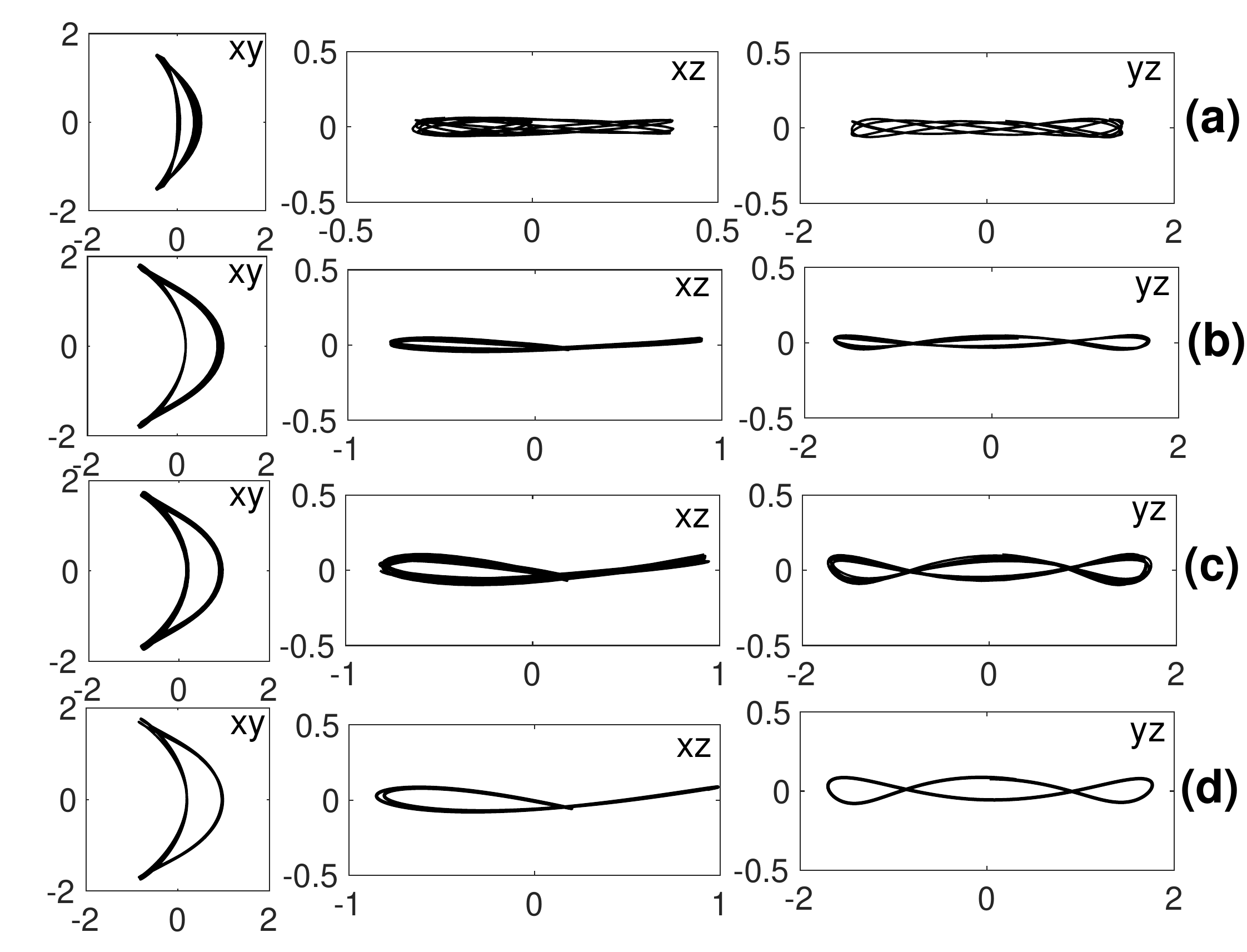}
 \caption{2D projections on the $(x,y)$ (left column), $(x,z)$ (middle column) and  $(y,z)$ (right column) planes of quasi-periodic orbits around the 2D thr$_1$ and 3D thr$_{z1}$ POs obtained by a $\Delta z = 5 \times 10^{-2 }$ perturbation. (a) and (b): the perturbed thr$_1$ orbit for $E_j = -0.38$ and $E_j = -0.338$, respectively. (c) and (d): the perturbed thr$_{z1}$ orbit for $E_j = -0.3346$ and $E_j = -0.33$, respectively.}
\label{fig:f10}
\end{figure*}

We now turn our attention to the morphologies supported in the system's configuration space $(x,y,z)$ by quasi-periodic orbits around the thr$_1$ and thr$_{z1}$ POs close to the bifurcation energy $E_B$. For energies $E_j < E_B$, the 2D family thr$_1$ is stable. All three 2D projections, i.e.~$(x,y)$, $(x,z)$ and $(y,z)$, of its orbits obtained by perturbing the thr$_1$ PO along the $z$ axis by $\Delta z = 5 \times 10^{-2 }$, for energies  $E_j = -0.38$ and  $E_j = -0.338$, are presented in Figs.~\ref{fig:f10}(a) and \ref{fig:f10}(b), respectively. For energies $E_j > E_B $ thr$_{z1}$, and its symmetric family thr$_{z1}$S, are stable. In Fig.~\ref{fig:f03}(c), we have already given the 2D projections of the 3D thr$_{z1}$ PO for $E_j = -0.3306$.  The quasi-periodic orbits obtained by a $\Delta z = 5 \times 10^{-2 }$ perturbation of the thr$_{z1}$ POs for $E_j = -0.3346$ and $E_j = -0.33$ are respectively shown in Figs.~\ref{fig:f10}(c) and \ref{fig:f10}(d). 

By comparing the projections of the 3D quasi-periodic orbits around the 2D stable thr$_1$  PO  for $E_j < E_B$ [Figs.~\ref{fig:f10}(a) and \ref{fig:f10}(b)] to the ones of the 3D stable thr$_{z1}$ PO [Fig.~\ref{fig:f03}(c)] and those of the quasi-periodic orbits for $E_j = -0.3346$ [Fig.~\ref{fig:f10}(c)] and $Ej = -0.33$ [Fig.~\ref{fig:f10}(d)], for which we have   $E_j > E_B$, we see that all orbits have similar morphologies on the $(x,y)$ plane. On the other hand, the  $(x,z)$  and  $(y,z)$ projections of the perturbed thr$_1$ orbit for energies further away from the bifurcation point [Fig.~\ref{fig:f10}(a)] are quite different from the ones exhibited by the 3D thr$_{z1}$ PO [Fig.~\ref{fig:f03}(c)] and its perturbation [Figs.~\ref{fig:f10}(c) and \ref{fig:f10}(d)]. Nevertheless, for energies closer to, but still below the bifurcation point $E_B$, the  $(x,z)$  and  $(y,z)$ projections of the 3D  perturbation of the 2D stable thr$_1$ PO [Fig.~\ref{fig:f10}(b)] become very similar to the ones shown by the thr$_{z1}$ PO [Fig.~\ref{fig:f03}(c)] and its perturbed orbits [Figs.~\ref{fig:f10}(c) and \ref{fig:f10}(d)], despite the fact that the thr$_{z1}$ family does not even exist for $E_j < E_B$. Therefore, we can infer that, to some extent, the morphology in the configuration space of the 3D perturbations of the 2D thr$_1$ POs close before the bifurcation ($E_j < E_B$) foreshadows the shape of the 3D PO which \textit{will be} introduced in the system at a large energy ($E_j = E_B$). This evolution of the shapes of the quasi-periodic orbits towards the morphology of the upcoming bifurcating family has already been noticed earlier by \citep{patsis2014phasea} in a different case.

\subsection{Two 3D period-doubling bifurcations} 
\label{Sec:(mul2)thrz1}

The 3D thr$_{z1}$ family has a transition from stability to simple instability at energy $E_C = -0.3203$. At this point its $b_2$ stability index becomes  $b_2>2$ (see Fig.~\ref{fig:f02}), and a period-doubling bifurcation takes place \citep[Sect.~2.11.2]{contopoulos2002order}, leading to the creation of the 3D family thr$_{z1}$(mul2) and its symmetric counterpart thr$_{z1}$(mul2)S. As we can observe in Fig.~\ref{fig:f02} the thr$_{z1}$(mul2) family of POs is initially stable, becoming unstable at $E_D=-0.2943$, having also an interval of double instability for $-0.2807 < E_j < -0.2692$. The morphology of this family is given in Fig.~\ref{fig:f03}(d), where its member for $E_j = -0.3183$ is depicted. 

At energy $E_D$, the multiplicity four, 3D family thr$_{z1}$(mul4) (magenta curves in Fig.~\ref{fig:f01}), together with its symmetric counterpart thr$_{z1}$(mul4)S are introduced in the system through a period-doubling bifurcation of the thr$_{z1}$(mul2) family. The thr$_{z1}$(mul4) family is initially stable and it becomes complex unstable in the energy intervals corresponding to the magenta shaded regions in Fig.~\ref{fig:f02}, namely for $-0.2917 < E_j < -0.2872$, and $-0.2667 < E_j < -0.2622$. A stable member of this family, for $E_j=-0.2831$, is depicted in Fig.~\ref{fig:f03}(e). 

In order to investigate the impact of these two period-doubling bifurcations on the phase space structure more deeply, we analyze four distinct cases with energies $E_i > E_C$, where $i = 1, 2, 3, 4$. These specific cases are selected to ensure that the four primary families of POs investigated in our study, namely thr$_1$, thr$_{z1}$, thr$_{z1}$(mul2), and thr$_{z1}$(mul4), exhibit diverse combinations of their kinds of stability. The $E_j$ values for these cases, the nature of stability exhibited by the specific POs, which are analyzed within the families, and the figures presenting the corresponding PSS are outlined in Table~\ref{tab:tab1}. We note that the values of $E_i$, $i=1, 2, 3, 4$ are indicated by vertical orange lines in Fig.~\ref{fig:f02}, with the first two being located between energies $E_C$ and $E_D$ [where the thr$_{z1}$(mul4)  family does not yet exist], and the other two being larger than $E_D$.
\begin{table*}[ht]
\tbl{\label{tab:tab1} In successive columns we give the energy values $E_i$, $i=1,2,3,4$, the stability kinds of the POs of the families thr$_1$, thr$_{z1}$, thr$_{z1}$(mul2), thr$_{z1}$(mul4) at these energies, and the figures' numbers where the corresponding PSS is presented, for the four specific cases discussed in Sect.~\ref{Sec:(mul2)thrz1}. S, U, DU and $\Delta$ respectively stand for stable, simple unstable, double unstable and complex unstable POs, while ``-'' denotes that the family does not exist at the specific energy.}
{\begin{tabular}{|c|c|c|c|c|c|c|}
\hline
Case & $E_j$ & thr$_1$ & thr$_{z1}$ & thr$_{z1}$(mul2)  & thr$_{z1}$(mul4)  & Figure\\
\hline
 1 & $E_1=-0.3183$ & U & U & S & - &\ref{fig:f11} \\ 
 2 & $E_2=-0.3157$ & DU & U & S & - &\ref{fig:f12}\\ 
 3 & $E_3=-0.2941$ & U & U & U & S &\ref{fig:f14} \\ 
 4 & $E_4=-0.2907$ & U & U & U & $\Delta$ &\ref{fig:f15} \\
\hline
\end{tabular}}
\end{table*}

\subsubsection{Case 1} 
\label{Sec:case 1}

\begin{figure*}[t]
\centering
\includegraphics[width=0.8\textwidth,keepaspectratio]{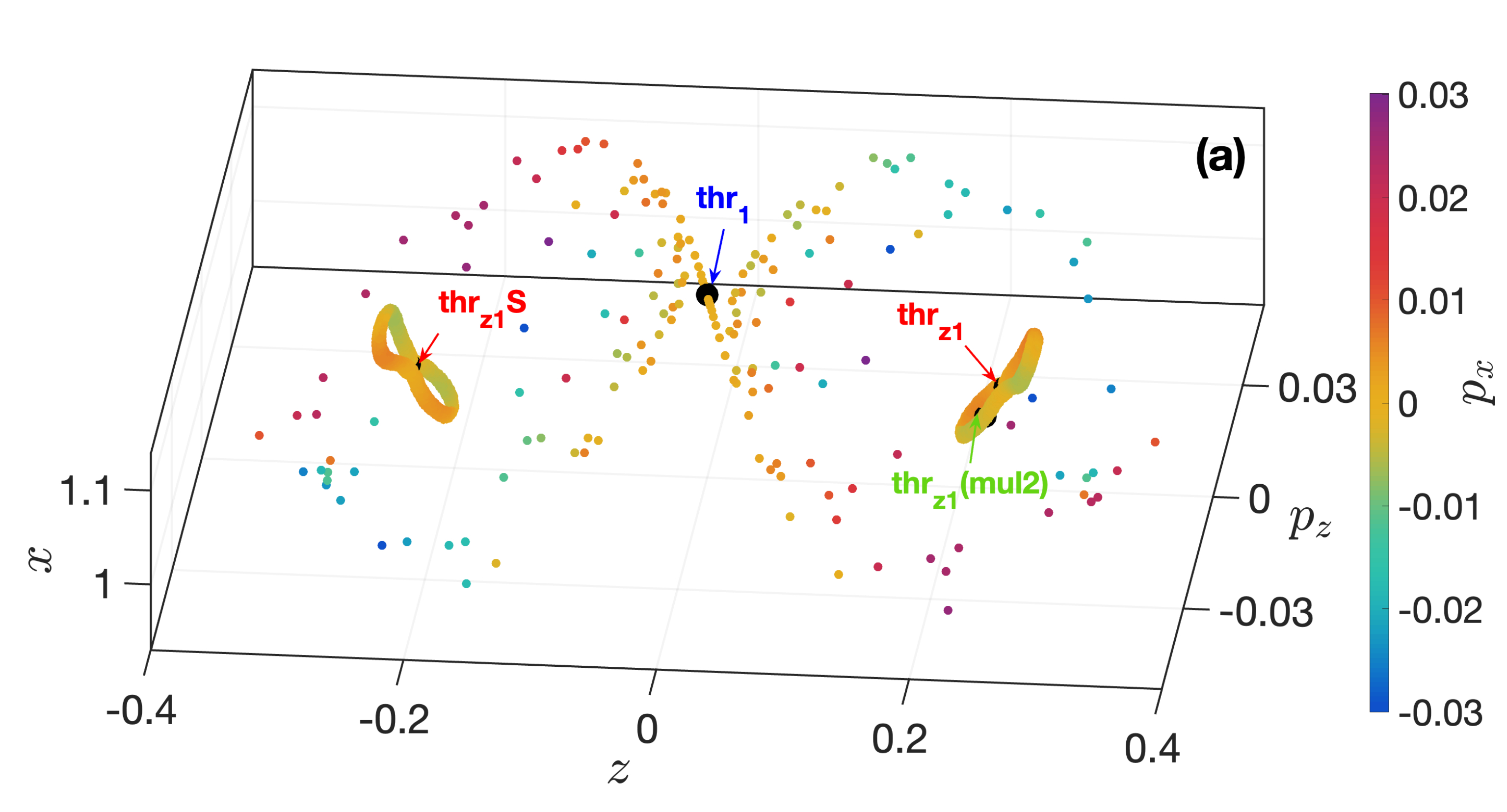}
\includegraphics[width=0.8\textwidth,keepaspectratio]{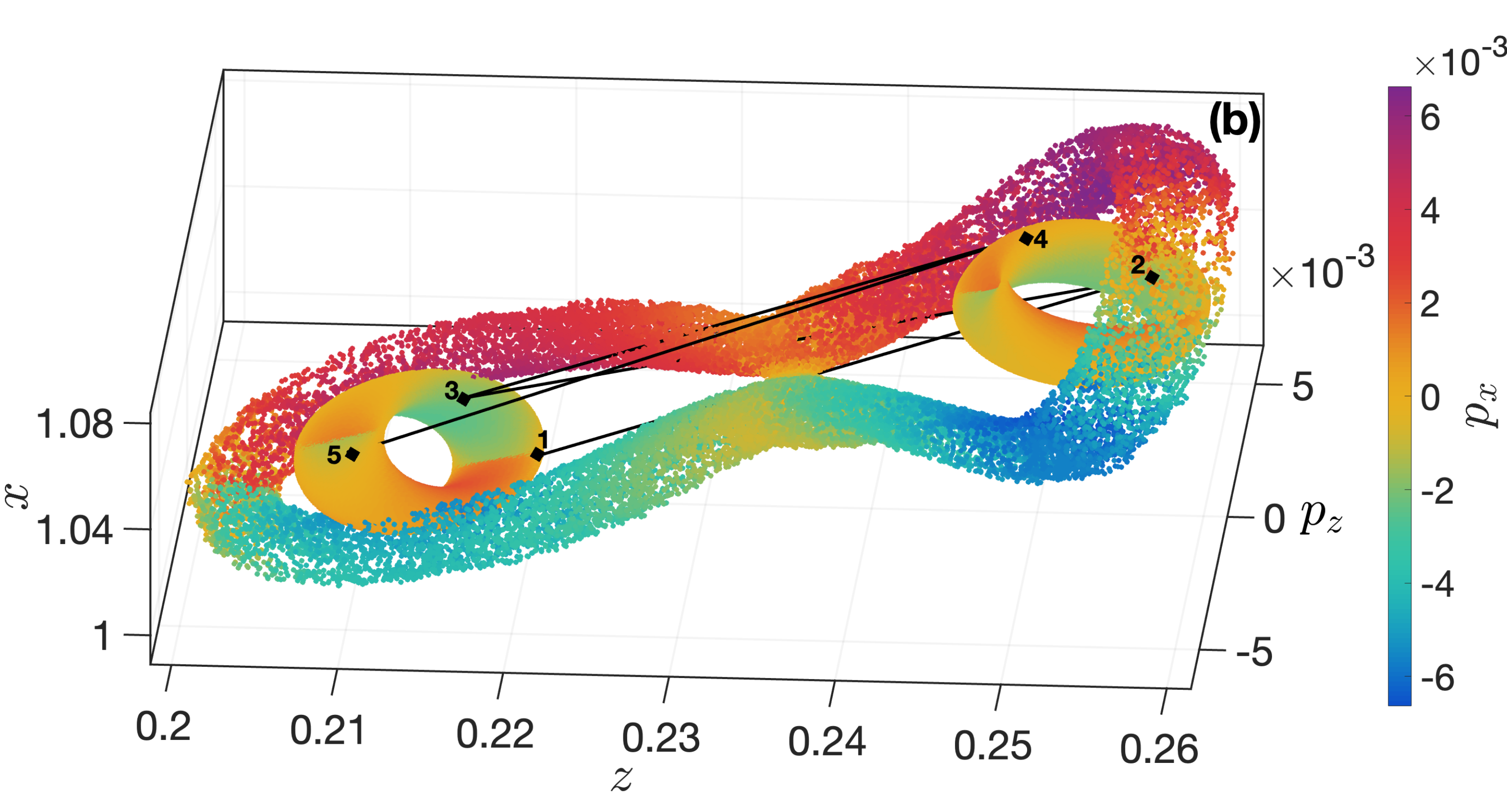}
 \caption{(Case 1 of Table \ref{tab:tab1}) The 3D colored $(x,z,p_z)$ projection of the system's 4D PSS at energy $E_1=-0.3183$. In (a) we observe orbits created by perturbations of the simple unstable thr$_1$ PO by $\Delta z = 10^{-5}$ (forming a loosely defined 8-shaped structure), and the simple unstable thr$_{z1}$ (right figure-8 structure) and thr$_{z1}$S (left figure-8 structure) POs by $\Delta z = 2.4\times 10^{-6}$. (b) A magnification of the figure-8 structure of panel (a), where also two tori created by a $\Delta z = 2.3\times 10^{-2}$ perturbation of the stable thr$_{z1}$(mul2) PO are included. The line-connected points denoted by black diamond symbols correspond to the first five consequents of a quasi-periodic orbit.}
\label{fig:f11}
\end{figure*}

In Fig.~\ref{fig:f02} we see that for energy $E_1=-0.3183$, just beyond the bifurcation point $E_C$,  the thr$_1$ (blue curves) and thr$_{z1}$ (red curves) POs are simple unstable, while the multiplicity two PO thr$_{z1}$(mul2) (green curves) is stable. As we can see in Fig.~\ref{fig:f11}(a), a $\Delta z=10^{-5}$ perturbation of the simple unstable thr$_1$ PO leads to a dispersed set of points, which vaguely form a figure-8 structure. This structure is not as well defined as in Fig.~\ref{fig:f07}. The thr$_{z1}$ PO, and its symmetric counterpart thr$_{z1}$S, are not stable anymore and thus they are not surrounded by tori, around which the thr$_1$ perturbations could form clear figure-8 structures. The existence of tori around the stable POs of the bifurcating families, during the transition of the parent family from stability to simple instability, plays a crucial role in sustaining the 8-shaped structure. As the consequents of the perturbed orbit are guided by the asymptotic curves of the unstable manifolds (cf.~Fig.~12 in \citep{katsanikas2013instabilities}), they give rise to   8-shaped structures sticking around the tori associated with the quasi-periodic orbits in the vicinity of the  stable POs of the bifurcating families.

Nevertheless, it is noteworthy that the scattered consequents of the perturbed thr$_1$ orbit in Fig.~\ref{fig:f11}(a) exhibit a smooth color variation, when they collectively shape the loosely defined 8-shaped structure. This suggests a sticky behavior, which may manifest within a significant time period for the dynamics of disk galaxies.

In any case, the perturbed by $\Delta z = 10^{-5}$ thr$_1$ orbit at $E_j = E_1$, exhibits different behavior compared to the thr$_1$ orbit perturbed by the same $\Delta z$ at $E_j = -0.3307$ (Fig.~\ref{fig:f07}). This serves as a notable illustration of the variations occurring in the vicinity of a PO (in this case, a simple unstable PO) as we move away from its bifurcation point. Therefore, whenever we refer to the typical phase space structures emerging  around unstable POs of various kinds, we are always referring to the orbits' immediate neighborhood as structured right after the bifurcation that introduces the PO into the system.

We also note that introducing a perturbation $\Delta x$ to the planar thr$_1$ orbit for $E_j = E_1$ leads to planar orbits. The nature of these orbits on the $(x, p_x)$ projection of the PSS depends on the magnitude of the perturbation; they may exhibit planar quasi-periodic behavior, or diffuse within the available phase space. In the specific case under consideration, we found orbits occupying the available phase space for perturbations with $\Delta x = 10^{-1}$, while for $\Delta x = 10^{-2}$, the orbit is quasi-periodic.

We observe a somewhat different outcome when we apply a perturbation $\Delta z = 2.4 \times 10^{-6}$ to the simple unstable thr$_{z1}$ (thr$_{z1}$S) PO at the right (left) side of Fig.~\ref{fig:f11}(a). This perturbation results in the formation of a figure-8 structure, well-defined this time, characterized by a smooth color variation. We note that the simple unstable thr$_{z1}$ (thr$_{z1}$S) PO resides at the junction of the two loops of the right (left) figure-8 structure in Fig.~\ref{fig:f11}(a). These two figure-8 formations are developed around tori corresponding to quasi-periodic orbits near the multiplicity two stable PO of the thr$_{z1}$(mul2) family [and the thr$_{z1}$(mul2)S]. Such  tori  created by a $\Delta z=2.3\times 10^{-2}$  perturbation of the thr$_{z1}$(mul2) PO  are given in  Fig.~\ref{fig:f11}(b). The figure-8 structure created by a $\Delta z= 2.4\times 10^{-6}$ perturbation of the simple unstable thr$_{z1}$ PO is also presented in the same panel. This arrangement is equivalent to the one seen in Fig.~\ref{fig:f07}, with the main difference being that the two tori in Fig.~\ref{fig:f11}(b) belong to the same orbit, while the ones depicted in Fig.~\ref{fig:f07} belong to two different orbits. The connection of the two tori in Fig.~\ref{fig:f11}(b) becomes apparent by tracing the location of the successive consequents starting by the initial condition identified by ``1'' and denoted by a black diamond on the left torus. By following the first five consequents of that orbit in Fig.~\ref{fig:f11}(b) (shown by line-connected black diamond symbols), we see that they alternatively belong to the left and the right torus. On the other hand, the consequents of an initial condition on the left or the right torus of  Fig.~\ref{fig:f07} remain on the same torus.

\subsubsection{Case 2} 
\label{Sec:case 2}

\begin{figure*}[t]
	\centering
	\includegraphics[width=0.8\textwidth,keepaspectratio]{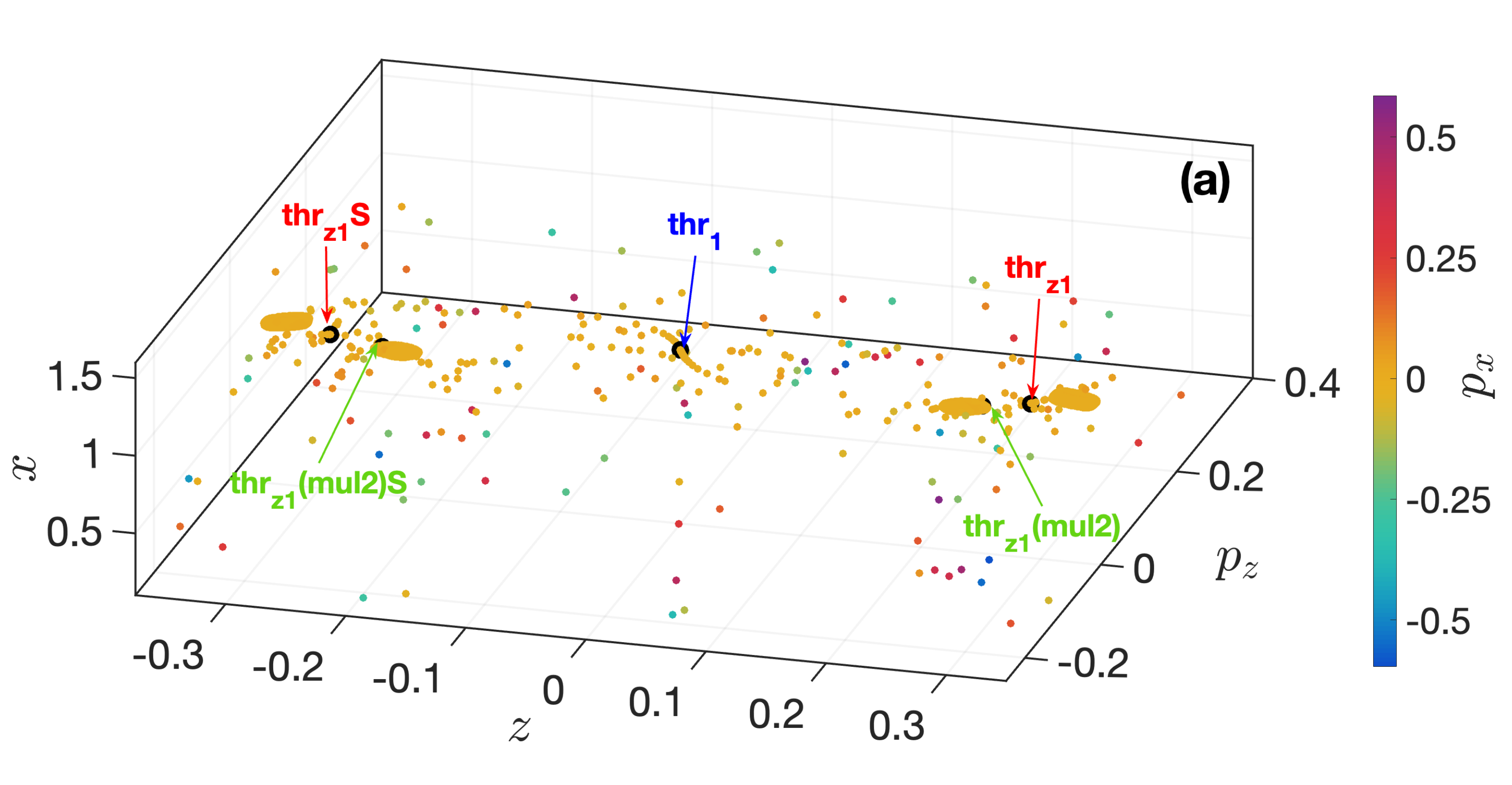}
	\includegraphics[width=0.8\textwidth,keepaspectratio]{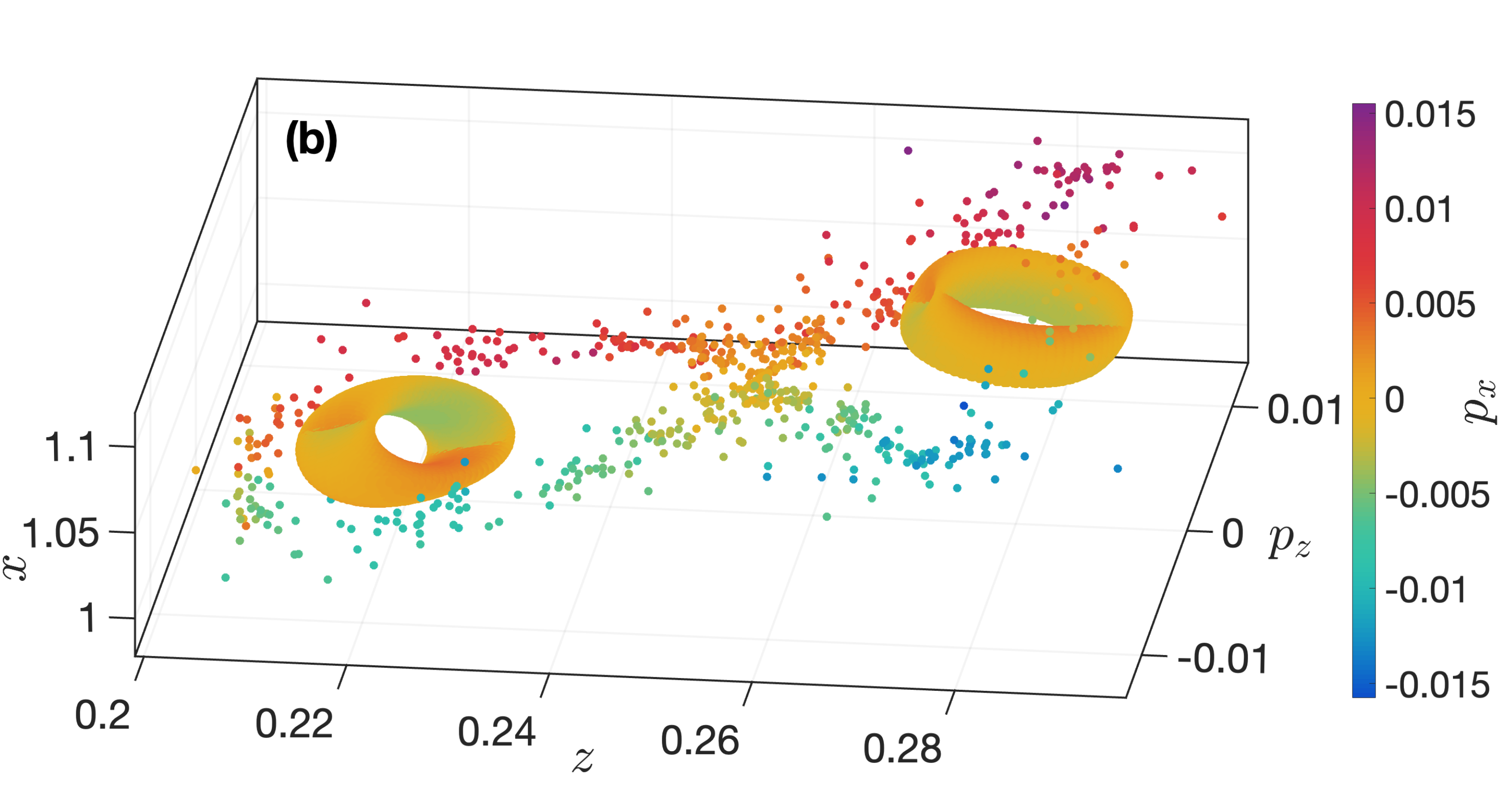}
   \caption{(Case 2 of Table \ref{tab:tab1}) The 3D colored $(x,z,p_z)$ projection of the system's 4D PSS at energy $E_2=-0.3157$. In (a) we observe the orbits created by perturbations of the double unstable thr$_1$ PO by $\Delta z = 10^{-5}$  (scattered points),  the simple unstable thr$_{z1}$ (right figure-8-like structure) and thr$_{z1}$S (left figure-8-like structure) POs by $\Delta z = 2.9\times 10^{-6}$. The figure also includes perturbed orbits of the stable thr$_{z1}$(mul2) (small yellowish tori for $z > 0$) and thr$_{z1}$(mul2)S (small yellowish tori for $z < 0$) by $\Delta z = 1.4\times 10^{-2}$ and $\Delta p_z = 5\times 10^{-3}$, respectively. (b) A zoomed-in region containing orbits around the thr$_{z1}$ (figure-8-like configuration) and thr$_{z1}$(mul2) (smooth tori) POs shown in the right part of panel (a) ($z>0$).}
\label{fig:f12}
\end{figure*}

We now move on to the case where $E_j=E_2$, \textit{just}  beyond the transition of thr$_1$ from simple to double instability. As we observe in Fig.~\ref{fig:f02}, at $E_2=-0.3157$, the thr$_1$ family (blue curves) is double unstable. In contrast, the thr$_{z1}$ and thr$_{z1}$(mul2) families (red and green curves respectively) retain the same instability they had in Case 1, i.e.~they respectively are simple unstable and stable. Now, for the thr$_1$ PO, a perturbation by $\Delta z = 10^{-5}$ of its initial condition leads to a very chaotic orbit whose consequents diffuse quickly, have mixed colors and look quite scattered in the 3D projection $(x,z,p_z)$ of the PSS [Fig.~\ref{fig:f12}(a)]. We do not discern any trace of a figure-8 formation, as in the vicinity of the member of the family presented in  Fig.~\ref{fig:f11}(a), where the thr$_1$  PO was simple unstable. 

The phase space structure in the neighborhood of the simple unstable PO thr$_{z1}$ at the same energy, $E_j=E_2$, is different. Perturbations by $\Delta z= 2.9\times 10^{-6}$ of this PO [right part of Fig.~\ref{fig:f12}(a), and Fig.~\ref{fig:f12}(b)] and its symmetric counterpart thr$_{z1}$S  [left part of Fig.~\ref{fig:f12}(a)] lead to the formation of an apparent figure-8-like structure, with a smooth color variation along it. This 8-shaped structure is not perpetual. Over time, the subsequent points of the orbit gradually distance themselves from this structure. What we are observing here is a form of sticky behavior. Nevertheless, these sticky figure-8 formations still surround smooth invariant tori created by perturbations ($\Delta z=1.4\times 10^{-2}$ and $\Delta p_z = 5\times 10^{-3}$) of the stable thr$_{z1}$(mul2) PO [Fig.~\ref{fig:f12}(b)]. By comparing Figs.~\ref{fig:f11}(b) and \ref{fig:f12}(b), we see that the figure-8 formation, created by a $\Delta z$ perturbation of the simple unstable thr$_{z1}$ PO becomes less well-shaped as energy increases. 

\begin{figure}[t]
	\centering
	\includegraphics[width=0.52\textwidth,keepaspectratio]{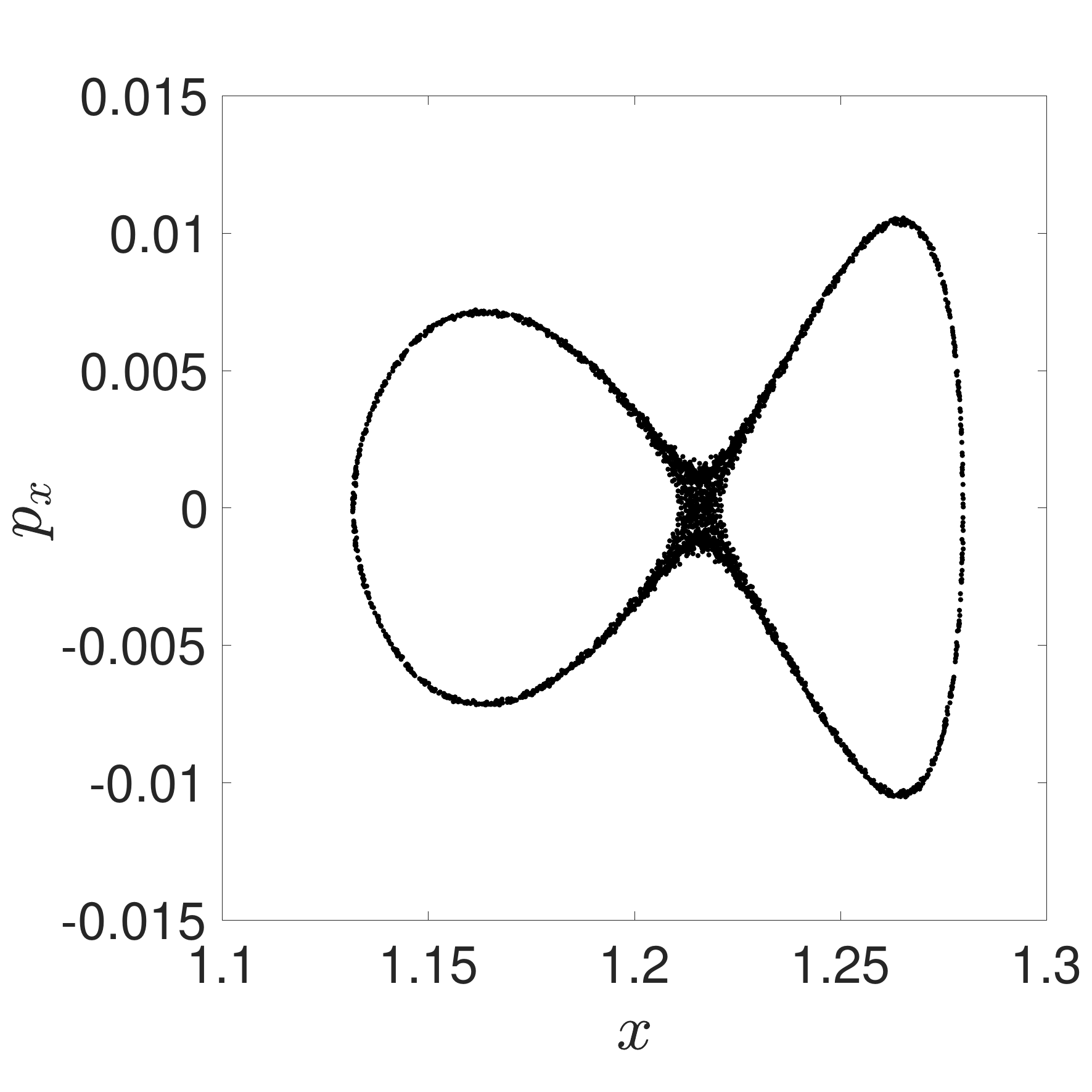}
 \caption{The $(x,p_x)$ PSS of a planar chaotic orbit obtained by perturbing the thr$_1$ PO by $\Delta x = 10^{-5 }$ at $E_2=-0.3157$.}
\label{fig:f13}
\end{figure}

Applying radial perturbations to POs at $E_j=E_2$ yields insightful outcomes. The planar family thr$_1$ exhibits double instability. The 4D PSS shows strong chaos in the vicinity of non-planar doubly unstable  POs. Typically, the consequents of the orbits form clouds of points with mixed colors \citep{katsanikas2013instabilities}. A $\Delta x$ perturbation of the thr$_1$ PO leads to a planar orbit. In the case of Fig.~\ref{fig:f13}, where the perturbation is $\Delta x = 10^{-5}$, the orbit does not display significant diffusion on the $(x,p_x)$ section. Instead, it forms a distinctive 8-shaped structure.

It is worth mentioning that $\Delta x$ perturbations of the thr$_{z1}$ family rapidly diffuse in phase space, following an initial period where they form an 8-shaped structure. This behavior is anticipated given the radially unstable nature of the simple unstable thr$_{z1}$ PO.

\subsubsection{Case 3} 
\label{Sec:case 3}

\begin{figure}[ht]
	\centering
       	\includegraphics[width=0.85\textwidth,keepaspectratio]{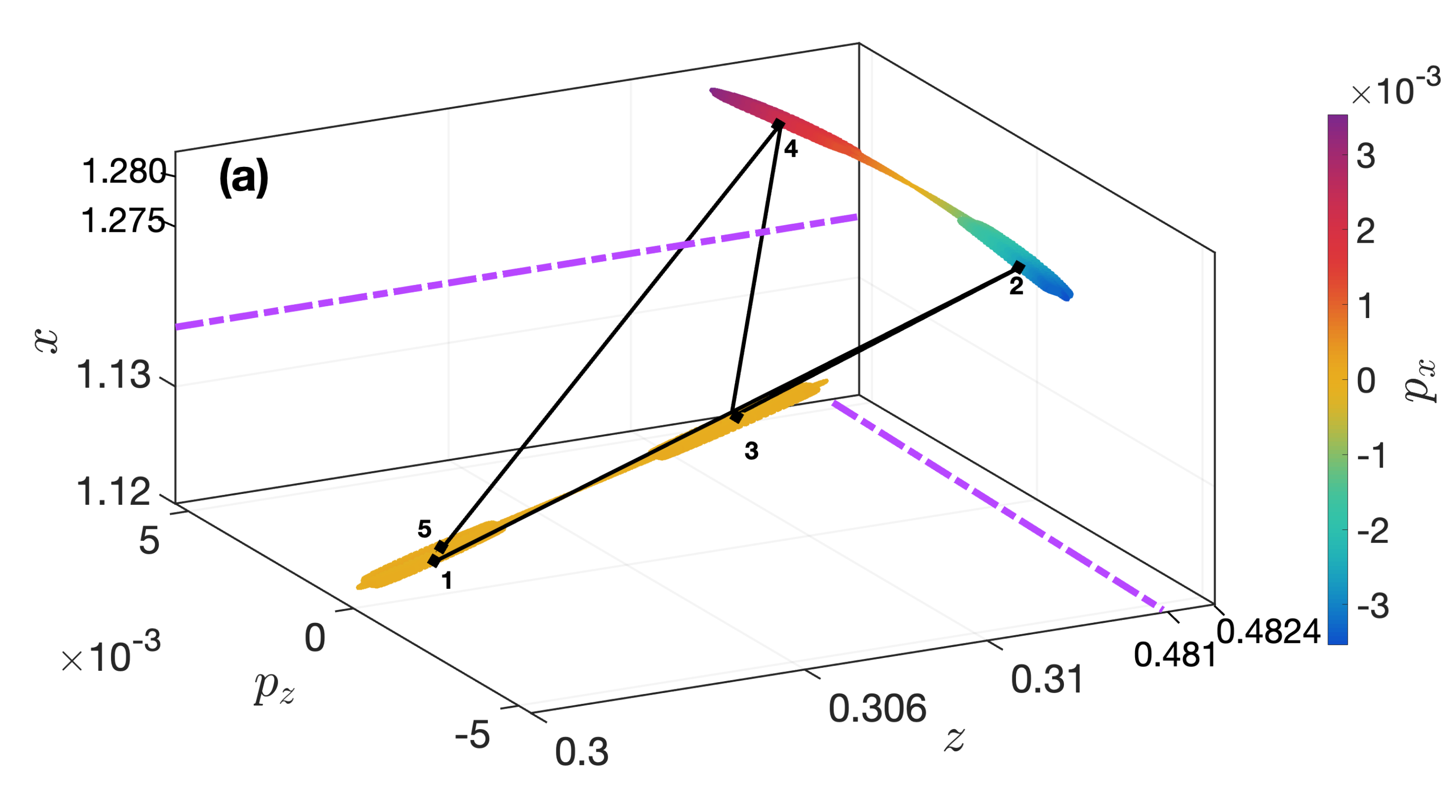}
  	\includegraphics[width=0.85\textwidth,keepaspectratio]{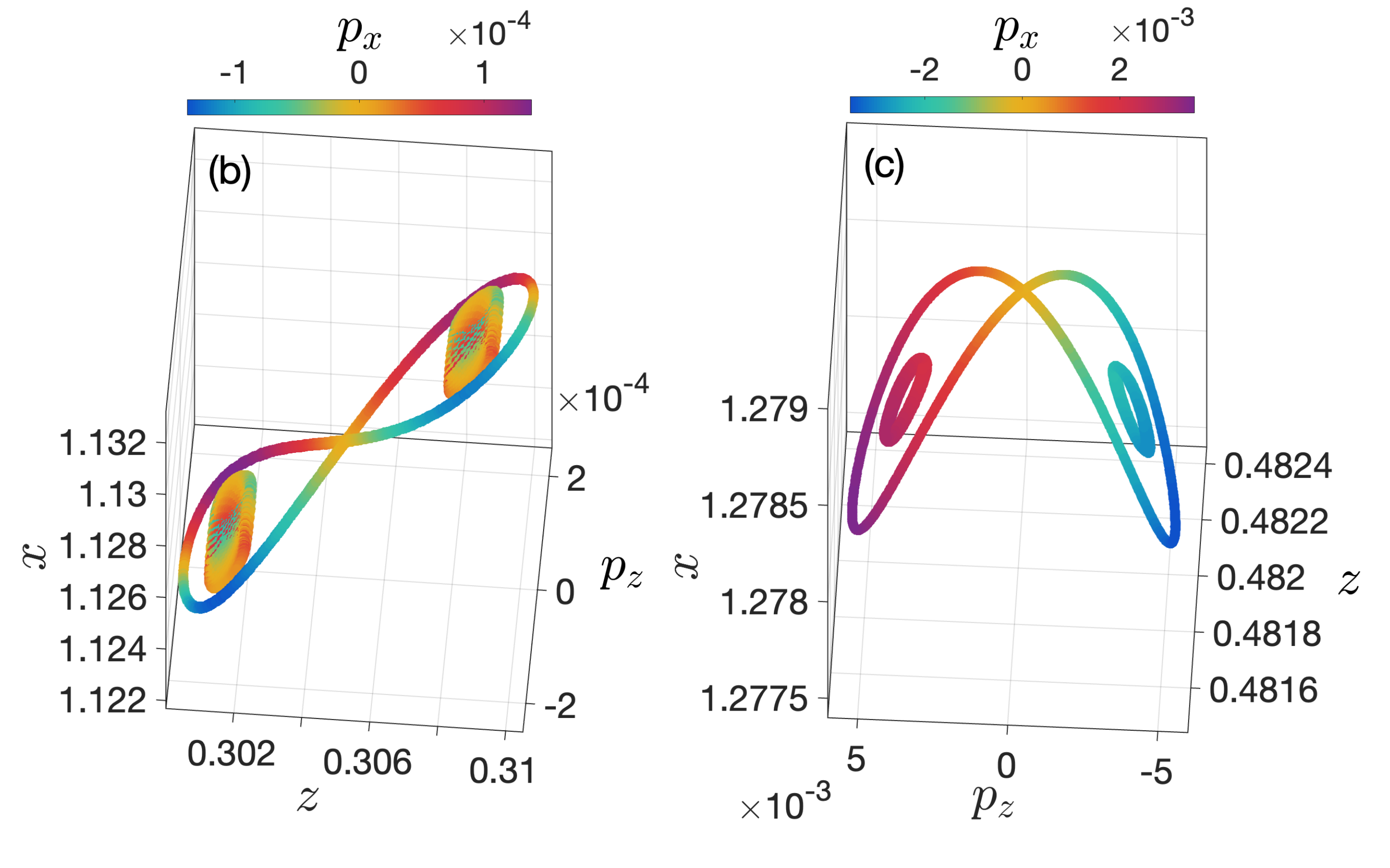} 
   \caption{(Case 3 of Table \ref{tab:tab1}) The 3D $(x,z,p_z)$ projection of the system's 4D PSS at energy $E_3=-0.2941$ with consequents colored according to their $p_x$ coordinate. A perturbation of the  multiplicity two, simple unstable thr$_{z1}$(mul2) PO ($\Delta z = 3\times 10^{-6}$) leads to the creation of two figure-8 structures,  while a perturbation of the multiplicity four, stable thr$_{z1}$(mul4) PO ($\Delta x = 10^{-5}$, $\Delta z = 6\times 10^{-5}$ and $\Delta p_z = 8\times 10^{-5}$) results to the creation of four tori. In (a), the $x$ and $z$ axes have been cropped around the purple dashed lines, while the numbered black, line-connected symbols show the first five consequents of an initial condition on one of the tori. (b) and (c) respectively show magnifications of the regions around the  lower and upper figure-8  structures in panel (a). }
\label{fig:f14}
\end{figure}

The next case we investigate is for $E_3=-0.2941$. At this energy, both thr$_1$ and thr$_{z1}$ POs are simple unstable and their $\Delta z$ perturbations lead to chaotic orbits, whose consequents create clouds of scattered points with mixed colors in every 3D projection of the system's PSS. On the other hand, a perturbation $\Delta z= 3\times 10^{-6}$ of the multiplicity two, simple unstable thr$_{z1}$(mul2) PO, having its radial stability index $b_2 > 0$, results in the formation of two figure-8 structures. They are formed in the vicinity of the two initial conditions of the PO [Fig.~\ref{fig:f14}(a)]. These two rather thin figure-8 formations reside around invariant tori in the vicinity of the stable PO thr$_{z1}$(mul4). We note that the thr$_{z1}$(mul4) PO is of multiplicity four, so its perturbations lead to the creation of four  tori, which are characterized by a smooth color variation on their surface. Such tori created by perturbing three coordinates  of the thr$_{z1}$(mul4) PO initial conditions ($\Delta x = 10^{-5}$, $\Delta z = 6\times 10^{-5}$ and $\Delta p_z = 8\times 10^{-5}$) are shown in Figs.~\ref{fig:f14} (b) and (c).  As the two figure-8 formations around the simple unstable thr$_{z1}$ (mul2) PO occur in phase space regions that are far apart from each other, we narrowed down the $x$ and $z$ axes displayed in Fig.~\ref{fig:f14}(a) to the area within the purple dashed lines (excluding the intervals $1.135 < x < 1.27$ and $0.315 < z < 0.48$). This adjustment was made to consolidate all structures into a single panel.

The black, line-connected symbols in Fig.~\ref{fig:f14}(a) depict the first five consequents of an initial condition on a torus around the stable thr$_{z1}$(mul4) PO, and clearly show the way the four tori are formed. Each one of the two figure-8 structures, along with the tori they surround, become more apparent in the magnifications provided in Fig.~\ref{fig:f14}(b), showing the tori where points ``1'', ``3'' and ``5'' are located, and  Fig.~\ref{fig:f14}(c) depicting the PSS region around the other two tori.  We also note that in each panel of Fig.~\ref{fig:f14}, we used a different color scale range to color the consequents according to their $p_x$ coordinate. This was done to distinctly showcase the smooth color variation across the different structures. 

Therefore, once more in this case, we observe a recurring pattern: the emergence of a figure-8 configuration near the parent simple unstable PO, encompassing tori around the bifurcated stable PO exhibiting a smooth color variation. This scenario mirrors what we witnessed in Fig.~\ref{fig:f07} following a pitchfork bifurcation, as well as in Figs.~\ref{fig:f11}(b) and \ref{fig:f12}(b), after a period-doubling one.

Regarding radial perturbations, it is notable that for $E_j = E_3$, even small $\Delta x$ perturbations (of the order of $\Delta x=10^{-8}$) applied to the planar simple unstable thr$_1$ PO result in a planar chaotic orbit. Similarly, minor perturbations of the 3D PO thr$_{z1}$ also lead to highly chaotic orbits, displaying clouds of scattered consequents with mixed colors in the 4D PSS we visualize using the color and rotation method. It is worth noting that, for both families, energy $E_3$ is significantly larger than the energies $E_A$ and $E_B$, at which these families were introduced in the system. Conversely, the recently bifurcated thr$_{z1}$(mul2) family, which is also simple unstable due to its radial instability, exhibits a similar behavior under radial perturbations at $E_3$ with what we observed when applying vertical perturbations. In other words, its perturbation leads to the appearance of persistent 8-shaped structures with smooth color variation.

\subsubsection{Case 4} 
\label{Sec:case 4}

\begin{figure}[ht]
	\centering
	\includegraphics[width=0.8\textwidth,keepaspectratio]{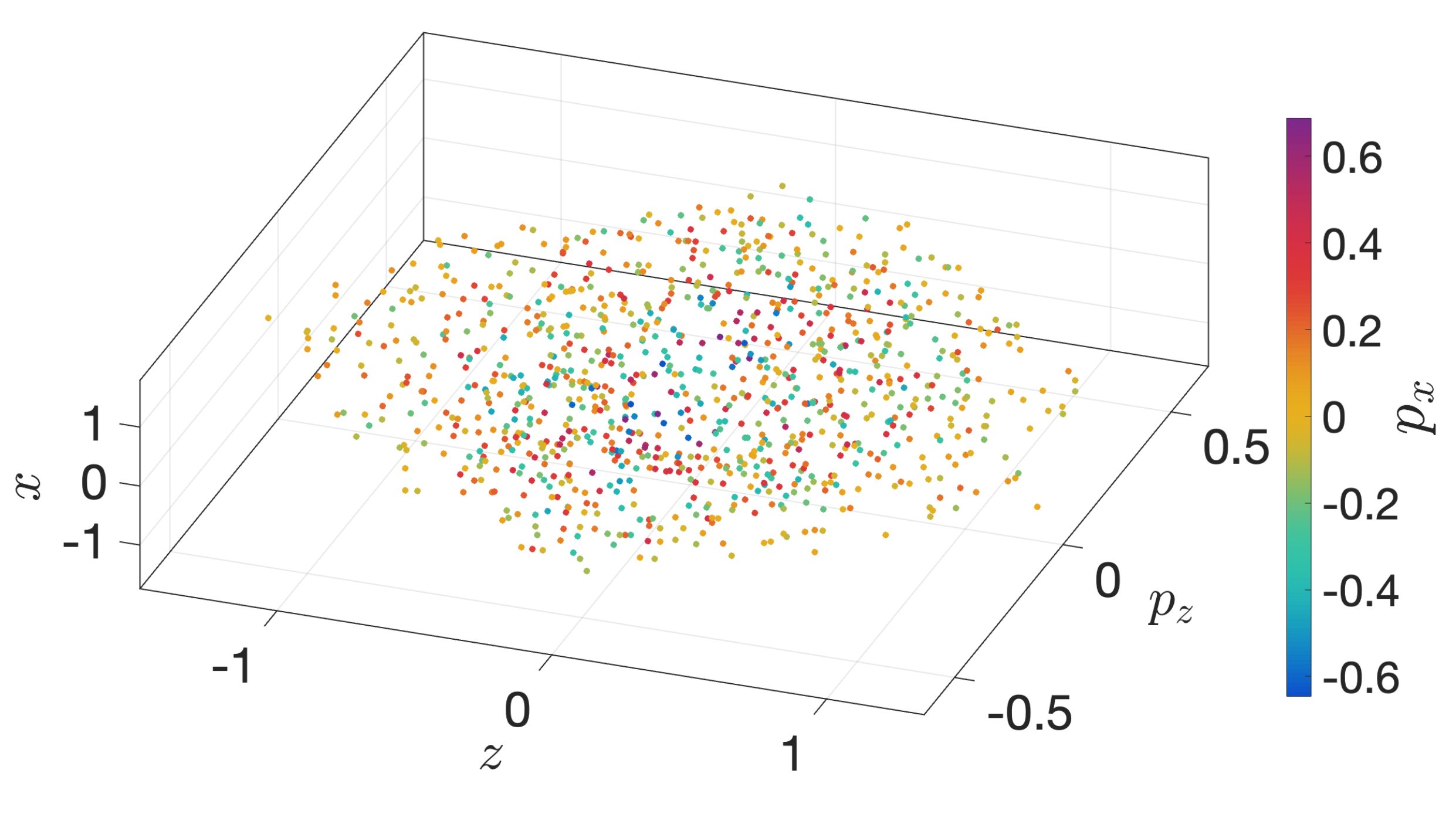}
   \caption{(Case 4 of Table \ref{tab:tab1}) The 3D colored $(x,z,p_z)$ projection of the system's 4D PSS at energy $E_4=-0.2907$ for a $\Delta z =10 ^{-5}$ perturbations of the simple unstable PO thr$_{z1}$.}
\label{fig:f15}
\end{figure}

The final case we consider is for energy $E_4=-0.2907$, for which the POs of all families are unstable. In particular, the thr$_1$, thr$_{z1}$ and  thr$_{z1}$(mul2) POs are simple unstable, while the thr$_{z1}$(mul4) PO is complex unstable. In this case, none of the considered POs have tori around them where regular motion occurs. The perturbation of any PO of the four families leads to chaotic orbits whose consequents form mixed-color clouds of scattered points, which eventually diffuse to larger regions of the PSS. Such an example is seen in  Fig.~\ref{fig:f15} where the 3D projection $(x,z,p_z)$ of the 4D PSS for a $\Delta z =10 ^{-5}$ perturbation of the thr$_{z1}$ PO is shown.
\section{Summary and discussion} 
\label{sec:Discussion and Conclusion}

We performed a detailed numerical investigation of the evolution of the phase space structure of the 3D galactic-type Hamiltonian system \eqref{eq:3DBarredGAlaxy} along a series of 2D and 3D bifurcations of families of POs. More specifically, we considered successively the pitchfork bifurcation of the planar 2D family thr$_{1}$  (and its symmetric counterpart thr$_{1}$S) from the 2D family x$1$ (Sect.~\ref{Sec:``8"}), and of the 3D family thr$_{z1}$ (and thr$_{z1}$S) bifurcated from thr$_{1}$ (Sect.~\ref{Sec:thrz1}). In these bifurcations, the parent family undergoes a transition from stability to simple instability, which results in the creation of two new families of stable POs of the same multiplicity. In addition,  we investigated in Sect.~\ref{Sec:(mul2)thrz1} how the phase space is structured after the 3D period-doubling bifurcations of thr$_{z1}$(mul2) from thr$_{z1}$ and of thr$_{z1}$(mul4) from thr$_{z1}$(mul2). In these latter cases, the transition of the parent family from stability to simple instability creates a child stable family of POs having twice the multiplicity of the parent one.  

In all studied cases, the consequents of orbits obtained by small perturbations of stable POs form well-defined tori in any 3D projection of the system's 4D PSS. These tori are characterized by a smooth variation of colors (colors indicate the value of the fourth PSS coordinate) across their surfaces. The variation of the colors is smooth but not necessarily simple. Across the tori, we observe a simultaneous color variation in both the poloidal and toroidal directions on the torus, along with a distinct shift of a primary hue from the inner to the outer side of the torus at specific regions (see e.g. Fig.~\ref{fig:f04}). This outcome, alongside earlier research \citep{katsanikas2011structure,zachilas2013structure}, suggests the ubiquity of this pattern concerning the phase space structure near stable POs in 3D autonomous Hamiltonian systems and 4D symplectic maps.

In the current study, we have delved deeper into examining how the phase space structure evolves in the neighborhood of a PO during its shift from stability to simple instability. This transition can manifest under different conditions:  After a pitchfork bifurcation we have the creation of two new families of stable POs sharing the same multiplicity with the parent family. Within the 4D PSS, these new stable POs are surrounded by two distinct sets of tori. This means that orbits originating from one torus consistently remain on the surface of the same torus (Fig.~\ref{fig:f07}). On the other hand, a period-doubling bifurcation results in the birth of families of POs having double the multiplicity of the parent family. In this case, perturbations of the created stable POs lead to the formation of more than one interconnected tori. For example, a regular orbit in the vicinity of the stable thr$_{z1}$(mul2) PO of multiplicity two forms two tori by having its consequents constantly alternating between them [Fig.~\ref{fig:f11}(b)]. In the same way, perturbations of multiplicity four, stable POs lead to the formation of four interconnected tori, as in Fig.~\ref{fig:f14}(a) for the case of the thr$_{z1}$(mul4) family.

In almost all cases of transitions from stability to simple instability (see also Fig.~\ref{fig:f13} for a case associated with double instability), a recurring pattern emerges: the appearance of figure-8 structures. These structures arise when perturbations are introduced to the parent family past the bifurcation point once it has become simple unstable. They encircle the tori formed from perturbations of the bifurcated stable POs and display a smooth color variation on their surfaces. An exception to this rule arises in the instance of planar simple unstable POs that are vertically unstable and radially stable, when subjected to radial perturbations. In such cases, there will always be a range of radial perturbations where the resulting planar orbits become quasi-periodic, as observed in the case of thr$_1$ just beyond $E = E_B$ (Sect.~\ref{Sec:thrz1}). Consequently, these orbits will manifest as invariant curves in the $(x,p_x)$ projection of the 4D PSS. 

The existence of the 8-shaped structures is connected to the existence of the tori of the regular quasi-periodic orbits around the stable bifurcated families. This holds in all kinds of bifurcations we examined, i.e.~the pitchfork bifurcation of family thr$_{z1}$ from thr$_{1}$, as well as the period-doubling bifurcations of families thr$_{z1}$(mul2) and thr$_{z1}$(mul4) from thr$_{z1}$ and thr$_{z1}$(mul2), respectively.

At energies distant from the bifurcation point, the coherence of the figure-8 structure diminishes, causing the points forming it to eventually disperse into phase space regions far from the unstable parent PO [Figs.~\ref{fig:f11}(a) and \ref{fig:f12}(b)]. We hardly observe such structures at energies where the tori of the bifurcated families are disrupted due to the instability of the parent POs. This emphasizes the correlation between the presence of the two structures (8-shaped figures and tori). Both persist as long as they are simultaneously present.

In general, the structure of phase space in the neighborhood of a PO is linked to the properties of the POs existing close by. As their properties vary with the energy, alterations take place in the initially formed structures. For this reason, we consider structures typical for a kind of orbital instability, as those observed just beyond the bifurcation, which introduces the related POs into the system.

An intriguing discovery from our research is associated with the shape of perturbed orbits near the 3D pitchfork bifurcation discussed in Sect.~\ref{Sec:thrz1}. As the system's energy approaches the bifurcation point (energy $E_B$ in this instance), the vertical perturbations of the 2D thr$_1$ PO [Fig.~\ref{fig:f10}(b)] generate structures resembling that of the bifurcated family thr$_{z1}$ [Figs.~\ref{fig:f10}(c) and \ref{fig:f10}(d)], even though this family has not yet been introduced into the system. In a way, the dynamical system anticipates the form of orbits that will arise following the bifurcation. This behavior has been observed previously (also referenced in \citep{patsis2014phasea}) and warrants further investigation regarding its implications for the study of analogous systems.

\nonumsection{Acknowledgments} 
H.~T.~M.~acknowledges funding from the Science Faculty PhD Fellowship  of the University of Cape Town, as well as  partial funding by the UCT Incoming International Student Award,  the Ethiopian Ministry of Education and the Woldia University. Ch.~S.~acknowledges support from the Research Committee (URC) of the University of Cape Town and the National Research Foundation (NRF) of South Africa. M.~H.~was supported by the Max Planck Society. We thank the High-Performance Computing facility of the University of Cape Town and the Center for High-Performance Computing (CHPC) of South Africa  for providing computational resources for this project. 

\appendix{3D perturbations of radially unstable and vertically stable POs} \label{AppA}

Let us make a note about the study of  3D  orbits near simple unstable POs that maintain vertical stability,  like the one depicted in Fig.~\ref{fig:f05}. In such cases it is crucial to employ a sufficiently small vertical  perturbation to unveil the phase space structure surrounding the PO. This precaution prevents the perturbed orbit from entering regions influenced by other nearby POs, such as the surface of a torus of a quasi-periodic orbit around a neighboring stable PO. Yet, with extremely minute perturbations, the ``width" of the torus might become significantly narrow, posing a challenge for standard graphics packages to distinguish between the front and back sides.

\begin{figure*}[t]
	\centering
 	\includegraphics[width=0.475\textwidth,keepaspectratio]{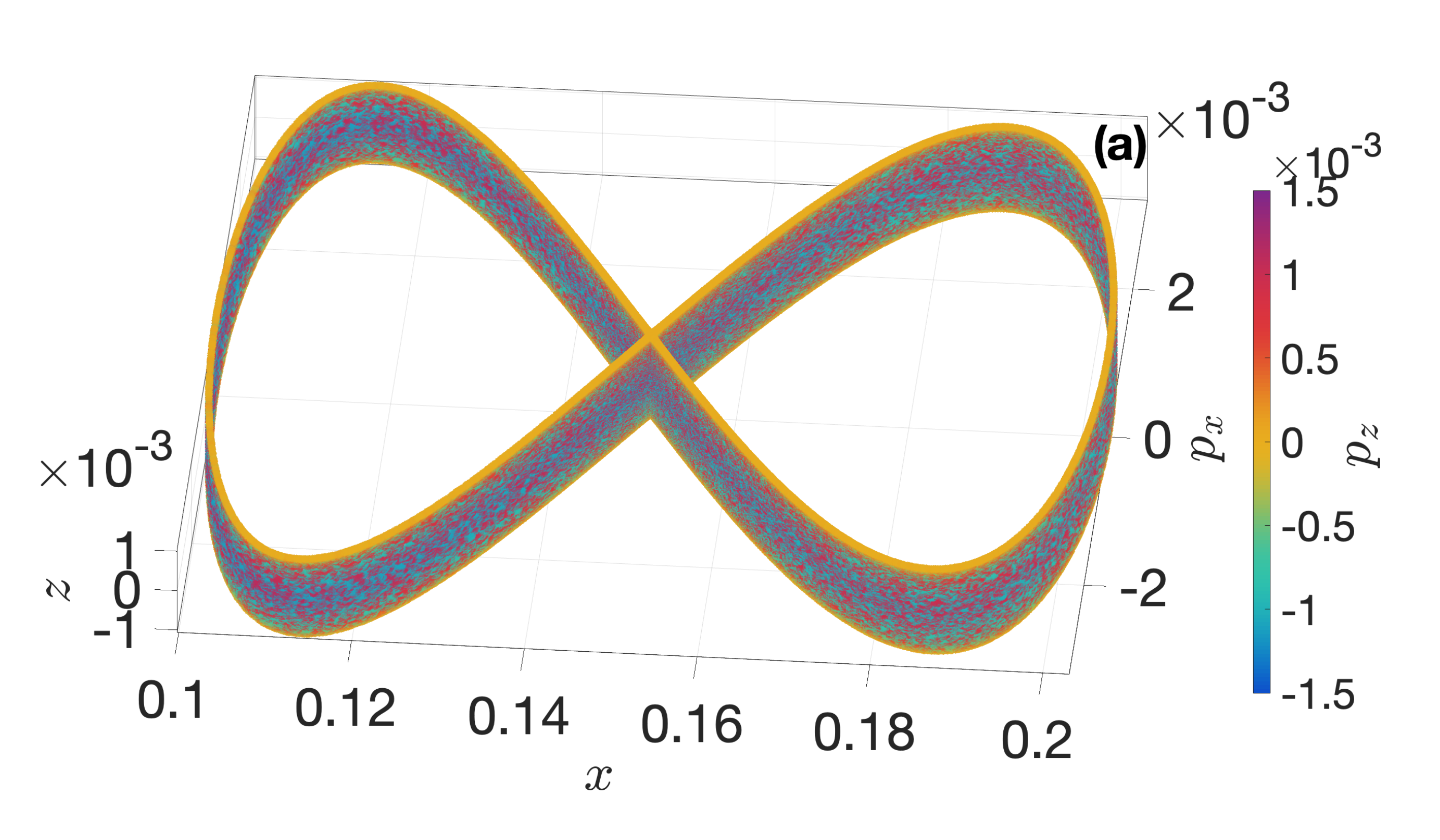}
	 \includegraphics[width=0.475\textwidth,keepaspectratio]{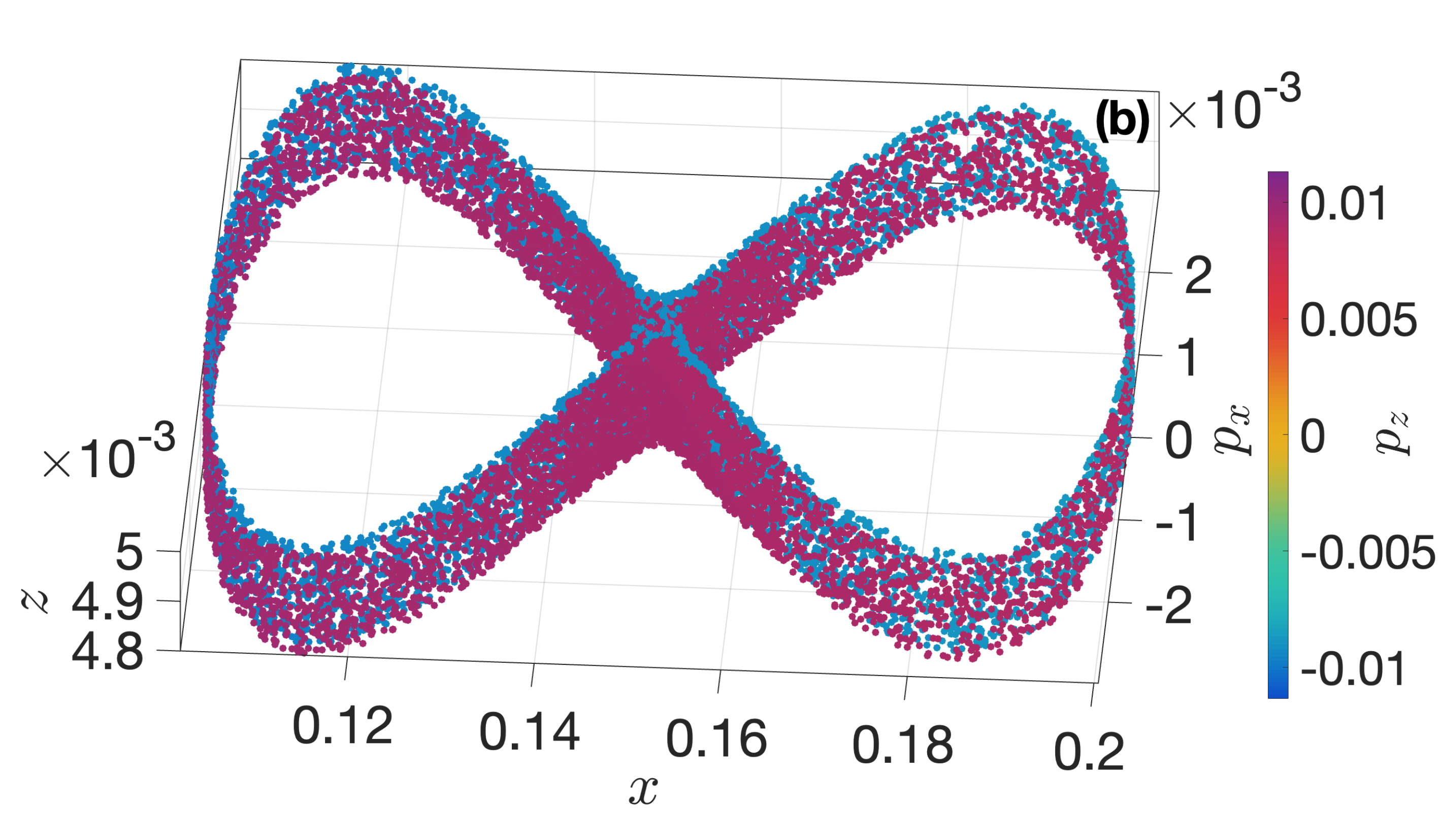}
	 \includegraphics[width=0.475\textwidth,keepaspectratio]{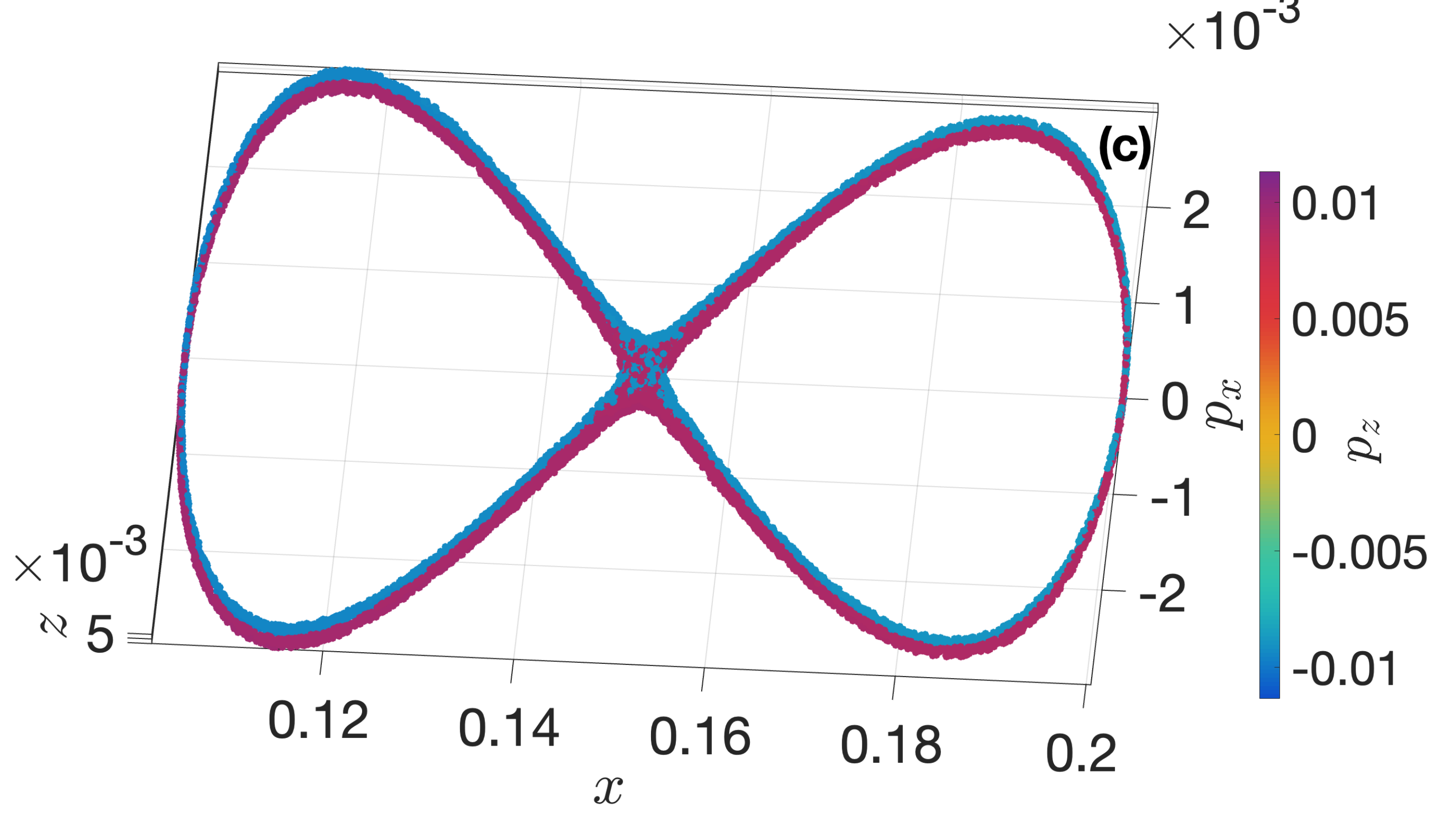}
	 \includegraphics[width=0.475\textwidth,keepaspectratio]{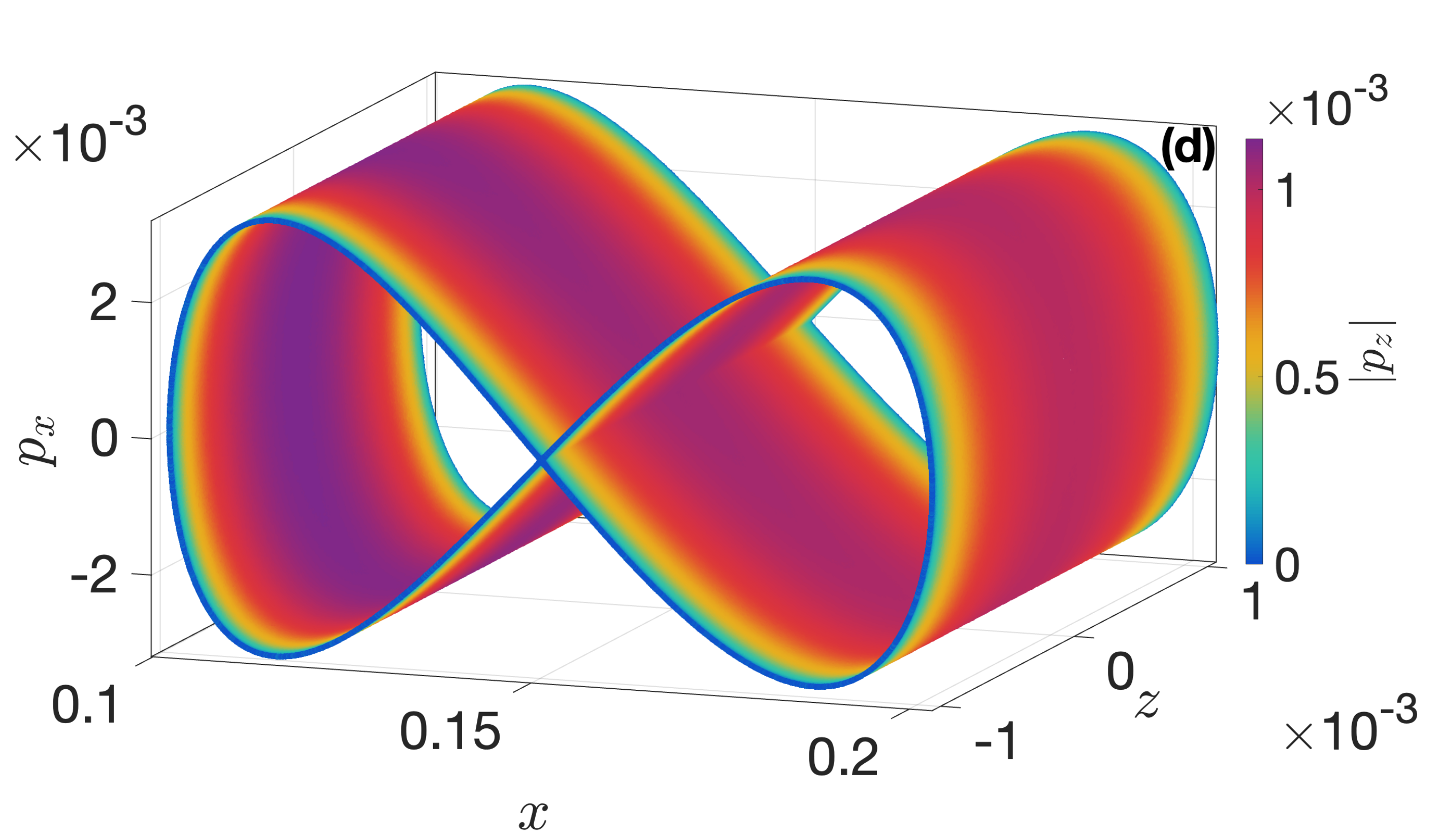}
	\caption{(a) The 3D colored $(x, p_x, z)$ projection of the system's 4D PSS for a perturbation of the unstable x$1$ PO by $\Delta z = \times 10^{-3}$ for $ E_j = -0.3919$. (b) Similar to (a), but for a narrower range of $z$ values $(0.0048 \leq z \leq 0.005)$. (c) The same data as in (a)  but for a different projection angle. Here it becomes apparent that blue and red hues do not mix. (d) The same data as in (a), but now points are colored  according to their $|p_z|$ value so that a smooth color variation becomes apparent.}
 \label{fig:f16}
\end{figure*}

In such instances, an additional surface of section could aid in verifying the smoothness of the color distribution. For example, the outcome of the prolonged integration of the perturbation by $\Delta z = 10^{-3}$ of the simple unstable x$1$ PO at the same $E_j$ value as in Fig.~\ref{fig:f05}, is shown in  Fig.~\ref{fig:f16}(a).  The apparent blending of colors here results from the narrow width of the ribbon-like structure in the $(x, p_x, z)$ projection. However, narrowing the range of the $z$ values even further,  for example by setting $0.0048 \leq z \leq 0.005$ [Fig.~\ref{fig:f16}(b)] and rotating the $(x, p_x, z)$ projection at an appropriate angle [Fig.~\ref{fig:f16}(c)], can lead to a clearer understanding of the created structure. In Fig.~\ref{fig:f16}(c) the colors segregate distinctly, even revealing the locations where they transition from one side to the other. In potentials having a symmetry with respect to a spatial variable, as in our case  where we have a symmetry with respect to the $z=0$ plane, a reliable indicator of smooth color variation can be obtained by coloring the consequents according to the absolute value of their fourth coordinate, i.e.~$|p_z|$ [Fig.~\ref{fig:f16}(d)]. 

\bibliographystyle{ws-ijbc}
\bibliography{IJBC_MKPHS}

\end{document}